\documentclass[12pt,a4paper]{article}

\usepackage[T1]{fontenc}
\usepackage{lmodern}
\usepackage[margin=25mm]{geometry}

\usepackage{authblk}

\usepackage{amsmath,amssymb,amsthm,mathtools,bm}

\usepackage{graphicx}
\usepackage{booktabs,tabularx,longtable,array,multirow}
\usepackage{siunitx}

\usepackage{xcolor}
\usepackage{xurl}
\usepackage{microtype}
\usepackage{enumitem}
\usepackage{placeins}

\usepackage[
backend=biber,
style=numeric,
sorting=none
]{biblatex}

\usepackage[
colorlinks=true,
citecolor=blue!55!black,
linkcolor=blue!55!black,
urlcolor=blue!55!black
]{hyperref}

\usepackage[nameinlink]{cleveref}
\usepackage{orcidlink}

\newtheorem{theorem}{Theorem}
\newtheorem{proposition}[theorem]{Proposition}
\newtheorem{corollary}[theorem]{Corollary}
\newtheorem{lemma}[theorem]{Lemma}

\newcommand{\eff}{\mathrm{eff}}
\newcommand{\dd}{\mathrm{d}}
\newcommand{\calD}{\mathcal{D}}
\newcommand{\avgP}{\mathcal{P}}

\newcommand{\Love}{\Lambda}
\newcommand{\Del}{\Delta}
\newcommand{\Order}{\mathcal{O}}

\renewcommand{\arraystretch}{1.16}
\title{Boundary Compatibility, Coupling--Compactness
	Locking, and Tidal Nonuniqueness in Metric-First
	Anisotropic Stars of Linear
	\texorpdfstring{$f(R,T)$}{f(R,T)} Gravity}

\author[1]{D. M. Pandya\,
	\orcidlink{0000-0002-2365-3587}%
	\thanks{Corresponding author: \href{mailto:dishantpandya777@gmail.com}{dishantpandya777@gmail.com}}}

\author[2]{Hemani R. Acharya}

\author[2]{Bharatkumar B. Parekh%
	\thanks{Corresponding author:
		\href{mailto:dr.bharatparekh@gmail.com}
		{dr.bharatparekh@gmail.com}}}

\author[3]{V. O. Thomas}

\affil[1]{Department of Mathematics,
	Pandit Deendayal Energy University,
	Gandhinagar 382426, Gujarat, India}

\affil[2]{Department of Physics,
	Pandit Deendayal Energy University,
	Gandhinagar 382426, Gujarat, India}

\affil[3]{Department of Mathematics,
	Faculty of Science,
	The Maharaja Sayajirao University of Baroda,
	Vadodara 390002, Gujarat, India}

\date{25 November 2026}
\begin{document}
	\maketitle
	
	\begin{abstract}
A positive surface density imposes a restriction on smooth stellar matching in linear $f(R,T)=R+2\chi T$ gravity. For nonzero coupling, the average-pressure prescription $L_m=-(p_r+2p_t)/3$ and the matter-variation relation used here imply $p_t(R)=3\rho(R)/2$ wherever the material and effective radial pressures vanish together. Such a surface violates the tangential dominant energy condition. For a general linear pressure-type matter Lagrangian we derive a geometry-independent condition on the effective radial-pressure slope. With average pressure, positive effective surface density and nonnegative effective radial pressure immediately inside the surface, this gives the necessary bound $M/R\geq 3/7$. When the material surface carries nonzero effective radial traction, continuous-metric matching to a Schwarzschild exterior requires an effective Israel layer with zero surface energy density and nonzero surface pressure. In a pseudo-spheroidal interior with prescribed effective radial pressure, the material boundary fixes the coupling $\chi$ as a function of the geometric and pressure-profile parameters. Four illustrative calibrations reproduce this relation. The quadrupolar tidal Love number also depends on how the bulk anisotropy and any surface layer respond to perturbations. Equilibrium geometry and static shell stress alone leave that response undetermined.
	\end{abstract}
	

	\section{Introduction}\label{sec:intro}
	Static and spherically symmetric stellar configurations have long provided an important setting for studying relativistic gravity. The Tolman--Oppenheimer--Volkoff equations \cite{Tolman1939,OppenheimerVolkoff1939}, Buchdahl's compactness theorem \cite{Buchdahl1959}, and Chandrasekhar's analysis of radial oscillations \cite{Chandrasekhar1964a,Chandrasekhar1964b} established an important distinction between three different questions: whether a static equilibrium solution can be constructed, whether the corresponding matter distribution is physically acceptable, and whether the equilibrium is dynamically stable. This distinction continues to be important in modern compact-star studies. Conditions such as regularity, positivity of the matter variables, subluminal characteristic speeds, hydrostatic force balance, and local adiabatic-index criteria provide useful checks on an equilibrium model, but none of them by itself replaces a well-defined perturbation analysis \cite{DelgatyLake1998,HerreraSantos1997,Abreu2007,Horvat2011,Pretel2021}.
	
	Pressure anisotropy introduces additional structure into the stellar matter. Differences between the radial and tangential pressures may arise from elastic stresses, magnetic fields, superfluid components, phase transitions, viscosity, composition gradients, or other forms of unresolved microphysics \cite{Ruderman1972,BowersLiang1974,HerreraSantos1997,DevGleiser2002,DevGleiser2003,Ivanov2002}. At equilibrium, the anisotropy may be written as
	\[
	p_t=p_r+\Delta.
	\]
	A perturbative calculation, however, also requires a prescription for how $\Delta$ changes away from equilibrium. Consequently, two models may have the same equilibrium profiles $\rho(r)$, $p_r(r)$, and $p_t(r)$ while having different radial oscillation spectra, nonradial modes, or tidal responses if their perturbative constitutive relations differ \cite{Horvat2011,BiswasBose2019,Lau2024,Guedes2026,Andersson2026}. This distinction becomes especially important when an exact equilibrium solution is used to calculate a tidal Love number.
	
	Exact and semi-analytic stellar interiors remain useful because they make regularity, scaling, parameter dependence, and boundary matching particularly transparent. Well-known examples include the Finch--Skea, Krori--Barua, Vaidya--Tikekar, Karmarkar-class, and pseudo-spheroidal constructions \cite{FinchSkea1989,KroriBarua1975,VaidyaTikekar1982,TikekarThomas1998,Karmarkar1948,Lake2003}. Their physical interpretation, however, requires a clear distinction between parameters belonging to the gravitational theory, parameters introduced through the matter or geometric prescription, integration constants, and quantities fixed from observational inputs. A collection of stellar configurations that individually satisfy the adopted conditions does not necessarily form a single stellar sequence, and changing only an overall length scale is not equivalent to varying the central density while keeping the same underlying microphysics.
	
	This distinction becomes more important in $f(R,T)$ gravity, where the gravitational action depends explicitly on the trace $T$ of the matter energy--momentum tensor \cite{Harko2011}. The resulting field equations depend not only on the chosen function $f(R,T)$, but also on the form of the matter Lagrangian and the associated tensor $\Theta_{\mu\nu}$. The role of the matter Lagrangian and the ambiguities associated with different fluid descriptions have been discussed from several viewpoints \cite{Brown1993,Faraoni2009,HarkoLobo2010,MinazzoliHarko2012,BarrientosRubilar2014,AvelinoAzevedo2018,FisherCarlson2019}. Different prescriptions such as $L_m=-p$, $L_m=-\rho$, $L_m=-p_r$, or $L_m=-(p_r+2p_t)/3$ therefore lead to different relations between the physical matter variables and the corresponding effective source. Results obtained using different matter-Lagrangian choices cannot be directly combined without rederiving the physical-to-effective mapping and the associated nonconservation relation.
	
	The linear model $f(R,T)=R+2\chi T$ has been widely applied to compact stars using equations of state, prescribed metric potentials, anisotropy models, electric charge, and embedding conditions \cite{Moraes2016,Das2016,Deb2018,Biswas2019AnnPhys,Dey2021,RejBhar2021,RejBharGovender2021,Pretel2021,Carvalho2021,Chowdhury2025,Nazar2026}. These studies show that the matter--geometry coupling can modify the equilibrium structure and mass--radius behaviour of stellar configurations. A methodological issue nevertheless arises in some metric-first constructions. Observed values of $M$ and $R$ may first be inserted into the matching conditions to determine a geometric parameter, after which the material surface condition is solved for the coupling $\chi$. Repeating the same procedure for different stellar objects then produces different values of a parameter that is intended to belong to the gravitational action. Such star-dependent values may be useful as phenomenological inverse calibrations, but they cannot be interpreted directly as independent measurements of one universal coupling without establishing their statistical consistency within a common model.
	
	The treatment of the stellar boundary introduces a second important issue. Smooth matching, distributional matching, and matching through a thin shell represent physically different boundary constructions \cite{Darmois1927,Israel1966,MarsSenovilla1993,Senovilla2013,Poisson2004,Rosa2021}. In the linear $f(R,T)$ model considered here, the exterior spacetime is Schwarzschild when the vacuum field equations reduce to those of general relativity. Continuity of the metric at the stellar surface, however, does not by itself guarantee continuity of the extrinsic curvature. If the effective radial traction remains nonzero at the material boundary, an additional surface stress is required. Such a surface layer also introduces its own perturbative degrees of freedom and therefore becomes relevant when the tidal response is considered.
	
	The tidal Love number provides a useful way to expose this distinction between the equilibrium background and the perturbative matter response. For an isotropic perfect fluid, the static quadrupolar problem can be expressed through an interior perturbation equation together with an exterior matching relation \cite{FlanaganHinderer2008,Hinderer2008,BinningtonPoisson2009,DamourNagar2009,Postnikov2010,Hinderer2010,DamourNagarVillain2012}. More general systems involving anisotropy, self-bound surfaces, elastic layers, rotation, or internal interfaces show that the logarithmic derivative of the perturbation variable need not be continuous across the stellar surface \cite{LandryPoisson2015,Pani2015,BiswasBose2019,LauLeungLin2019,TakatsyKovacs2020,Gittins2020,ArangurenVera2024}. Any jump must instead follow from the properties assigned to the matter or interface. Thus, applying an isotropic effective-fluid perturbation equation to an anisotropic configuration without specifying how $p_t-p_r$ responds to the perturbation does not define a unique tidal problem.
	
	The present work brings the boundary-compatibility and tidal-closure questions into a common framework. We first derive the condition required for the material and effective-fluid surfaces to coincide at finite density for a general linear pressure-type matter Lagrangian. This leads to a geometry-independent lower compactness condition and to explicit results for commonly used choices of $L_m$. We then obtain the surface stress required when the metric remains continuous but the effective radial traction does not vanish. The general analysis is subsequently applied to a pseudo-spheroidal stellar model with a prescribed effective radial-pressure profile, for which the material boundary condition gives an exact relation between the gravitational coupling and the compactness-dependent model parameters. At fixed universal $\chi$, the same relation restricts the allowed compactness classes, while varying only the dimensional scale produces a homologous family rather than an equation-of-state-generated stellar sequence. Finally, we examine the tidal problem and show that neither the equilibrium background nor the background surface stress is sufficient to determine a unique Love number without additional information describing the perturbative response of the bulk anisotropy and, where present, the surface layer.
	
	The purpose of this analysis is not to rule out pseudo-spheroidal geometry itself. Rather, the restrictions identified here arise from the combined choice of metric ansatz, effective radial-pressure profile, scale relation, fixed profile parameters, and surface prescription. Separating these assumptions from the underlying geometry is therefore essential for interpreting the resulting stellar configurations and their tidal response.
	
	\subsection{Position relative to the existing literature}
	
	Three areas of the literature are directly relevant to the present problem, although they have largely developed along separate lines. The first concerns exact and semi-analytic stellar models, where one or more metric functions, matter profiles, embedding conditions, or equations of state are specified and the remaining field equations are solved accordingly \cite{DelgatyLake1998,Lake2003,VaidyaTikekar1982,TikekarThomas1998,FinchSkea1989,KroriBarua1975,Karmarkar1948}. Such constructions are valuable because they allow the geometry, scaling behaviour, and boundary conditions to be examined analytically. At the same time, care is required when interpreting the role of the model parameters, since boundary conditions may determine quantities that were initially introduced with a different physical meaning.
	
	The second area is compact-star modelling in trace-coupled gravity. In these theories, the effective source depends explicitly on the adopted matter Lagrangian, and different choices of $L_m$ lead to different physical-to-effective relations and nonconservation laws \cite{Harko2011,HarkoLobo2010,BarrientosRubilar2014,AvelinoAzevedo2018,FisherCarlson2019}. The third area concerns relativistic response theory, where anisotropy, density discontinuities, elastic layers, self-bound surfaces, and distributional interfaces can modify the matching conditions for perturbations \cite{Postnikov2010,LandryPoisson2015,Pani2015,BiswasBose2019,LauLeungLin2019,TakatsyKovacs2020,Gittins2020,ArangurenVera2024,Guedes2026,Andersson2026}.
	
	The present analysis connects these three lines of work through the stellar boundary-value problem. It does not introduce a new pseudo-spheroidal geometry, nor does it imply that metric-first constructions are generally inconsistent. Instead, the central questions are more specific: under what conditions can the material boundary determine a coupling parameter of the gravitational action, when does the same background require a surface layer, and what additional constitutive information is needed before a tidal response coefficient can be defined? Table~\ref{tab:literaturegap} summarizes how these questions relate to representative approaches in the literature. The main contribution of the present work is therefore the connection between these boundary, coupling, and response issues within one consistent framework, rather than a claim of priority for each individual ingredient.
	
	\begin{table}[t]
		\centering
		\caption{Relation of the present analysis to representative
			approaches. ``Background'' denotes an equilibrium solution,
			while ``response closure'' refers to a law describing the
			perturbation of anisotropic or surface stresses.}
		\label{tab:literaturegap}
		
		\small
		\setlength{\tabcolsep}{4pt}
		\renewcommand{\arraystretch}{1.20}
		
		\begin{tabularx}{\textwidth}{
				@{}
				>{\hsize=0.85\hsize\raggedright\arraybackslash}X
				>{\hsize=1.00\hsize\raggedright\arraybackslash}X
				>{\hsize=1.00\hsize\raggedright\arraybackslash}X
				>{\hsize=1.15\hsize\raggedright\arraybackslash}X
				@{}
			}
			\toprule
			
			\textbf{Approach}
			&
			\textbf{Primary input}
			&
			\textbf{Typical boundary treatment}
			&
			\textbf{Question addressed in the present work}
			\\
			\midrule
			
			Metric-first exact interiors
			
			\cite{FinchSkea1989,VaidyaTikekar1982,
				TikekarThomas1998,KroriBarua1975,Lake2003}
			&
			Metric potential(s), matter-profile ansatz,
			or embedding relation
			&
			Model constants are fixed using an adopted
			$(M,R)$, together with $p_r(R)=0$
			&
			Does the material surface leave the theory
			coupling independent, or does it determine
			it through the compactness?
			\\
			\addlinespace[6pt]
			
			Linear $f(R,T)$ stellar models
			
			\cite{Moraes2016,Das2016,Deb2018,
				Dey2021,RejBhar2021,Chowdhury2025,
				Nazar2026,Yusmantoro2026}
			&
			Gravity coupling, matter Lagrangian,
			equation of state, or exact geometry
			&
			Usually matched to a Schwarzschild exterior
			&
			Do the material and effective radial tractions
			vanish at the same surface for the chosen
			$L_m$?
			\\
			\addlinespace[6pt]
			
			Thin-shell and general junction theory
			
			\cite{Darmois1927,Israel1966,
				MarsSenovilla1993,Senovilla2013,
				Poisson2004,Rosa2021}
			&
			Interior and exterior geometries together
			with hypersurface data
			&
			Surface stress is determined by the jump
			in extrinsic curvature
			&
			What surface layer is required when the
			metric is continuous but
			$p_{r,\eff}(R)\neq 0$?
			\\
			\addlinespace[6pt]
			
			Relativistic tidal response
			
			\cite{Hinderer2008,BinningtonPoisson2009,
				LandryPoisson2015,BiswasBose2019,
				ArangurenVera2024,Guedes2026,
				Andersson2026}
			&
			Equilibrium background together with a
			perturbative constitutive law
			&
			Interior, exterior, and interface
			perturbations are matched consistently
			&
			What bulk and surface-response information
			is required beyond the equilibrium background
			to determine the Love number?
			\\
			
			\bottomrule
		\end{tabularx}
	\end{table}
	
	\subsection{Logical structure and scope}
	
	The analysis is arranged so that the assumptions behind each result remain clearly distinguished. The geometry-independent result of Sec.~\ref{sec:smoothsurface} requires only an Einstein-like effective source, a Schwarzschild exterior, finite surface density, and a smooth boundary shared by the material and effective-fluid descriptions. The thin-shell treatment in Sec.~\ref{sec:shell} keeps the metric continuous across the surface while allowing a discontinuity in the extrinsic curvature. The pseudo-spheroidal geometry and the prescribed effective radial-pressure profile are introduced only in Sec.~\ref{sec:pseudo}. The tidal analysis of Sec.~\ref{sec:tides} then addresses the perturbative response and does not assume a specific physical model for the surface layer.
	
	The conclusions should therefore be understood within the assumptions used to derive them. A dilute surface removes the finite-density restriction, a different choice of matter scalar changes the surface ratio $\kappa_s$, and allowing $R/L$ to vary modifies the corresponding compatibility relations. Similarly, specifying the constitutive response of both the bulk matter and the surface layer closes the tidal problem. This separation makes clear which results follow from the general boundary conditions and which depend on the particular closure adopted for the stellar model.

	\subsection{Assumptions and conventions}
	\label{sec:assumptions}
	We consider the linear matter--geometry coupling model
	\begin{equation}
		f(R,T)=R+2\chi T,
	\end{equation}
	where $\chi$ is a constant coupling parameter and
	$T=g^{\mu\nu}T_{\mu\nu}$ is the trace of the physical
	energy--momentum tensor. Throughout this work, we use
	geometrized units $G=c=1$ and the metric signature
	$(+,-,-,-)$.
	
	For a static, spherically symmetric configuration,
	the spacetime metric is written as
	\begin{equation}
		ds^2=e^{\nu(r)}dt^2-e^{\lambda(r)}dr^2
		-r^2(d\theta^2+\sin^2\theta\,d\phi^2).
	\end{equation}
	
	The anisotropic matter source is described by
	\begin{equation}
		T^\mu{}_\nu
		=
		\operatorname{diag}
		(\rho,-p_r,-p_t,-p_t),
	\end{equation}
	where $\rho$ is the energy density, $p_r$ is the
	radial pressure, and $p_t$ is the tangential pressure.
	Consequently,
	\begin{equation}
		T=\rho-p_r-2p_t.
	\end{equation}
	
	We adopt the general linear pressure-type matter scalar
	\begin{equation}
		\mathcal P
		=
		q_\rho\rho+q_rp_r+q_tp_t,
		\qquad
		L_m=-\mathcal P,
	\end{equation}
	where $q_\rho$, $q_r$, and $q_t$ are prescribed
	dimensionless coefficients.
	
	The variation of the physical energy--momentum tensor
	with respect to the metric introduces
	\begin{equation}
		\Theta_{\mu\nu}
		\equiv
		g^{\alpha\beta}
		\frac{\delta T_{\alpha\beta}}
		{\delta g^{\mu\nu}}.
	\end{equation}
	
	For the present phenomenological treatment, we adopt
	the matter-variation prescription
	\begin{equation}
		\Theta_{\mu\nu}
		=
		-2T_{\mu\nu}
		-\mathcal P g_{\mu\nu}.
		\label{eq:theta_prescription}
	\end{equation}
	
	This prescription is an explicit constitutive assumption.
	In general, specifying $L_m=-\mathcal P$ alone does not
	establish that all additional metric-variation terms
	vanish for an arbitrary anisotropic microscopic matter
	action. Accordingly, the physical-to-effective
	transformation and the boundary results derived below
	are conditional on Eq.~\eqref{eq:theta_prescription}.
	
	With this prescription, the field equations of the
	linear $f(R,T)$ model take the form
	\begin{equation}
		\boxed{
			G_{\mu\nu}
			=
			(8\pi+2\chi)T_{\mu\nu}
			+
			\chi(T+2\mathcal P)g_{\mu\nu}
		}.
		\label{eq:field}
	\end{equation}
	
	These equations can be written in the Einstein-like form
	\begin{equation}
		G_{\mu\nu}
		=
		8\pi T_{\mu\nu}^{\mathrm{eff}},
	\end{equation}
	where
	\begin{equation}
		8\pi T_{\mu\nu}^{\mathrm{eff}}
		=
		(8\pi+2\chi)T_{\mu\nu}
		+
		\chi(T+2\mathcal P)g_{\mu\nu}.
		\label{eq:effective_source}
	\end{equation}
	
	The effective energy density and principal pressures
	are therefore determined algebraically by the physical
	matter variables and the adopted matter scalar.
	For $\chi=0$, the equations reduce to those of
	general relativity.
	
	Since $f_R=1$ and the matter source vanishes outside
	the star, the vacuum field equations reduce to
	$G_{\mu\nu}=0$. We therefore adopt a Schwarzschild
	exterior for the configurations considered here.
	
	The stellar surface is treated according to the
	appropriate boundary class. Smooth matching requires
	continuity of the first and second fundamental forms,
	whereas a discontinuity in the extrinsic curvature
	is represented by an effective Israel surface layer.
	The corresponding surface stress and its perturbative
	response are distinguished from the bulk matter
	variables.
	
	Because the results depend on the metric signature,
	matter-variation prescription, and surface matching
	conditions, Table~\ref{tab:assumptions} summarizes
	the principal assumptions before the formal analysis.
	These conventions must be respected when comparing
	results obtained using different matter Lagrangians
	or metric signatures.

	\begin{table}[t]
		\caption{Assumptions and conventions used in the formal results. Items identified as specializations are introduced only in the corresponding later sections and are not required for the geometry-independent results.}
		\label{tab:assumptions}
		\centering
		\begin{tabularx}{0.96\textwidth}{
				p{0.22\textwidth}
				p{0.27\textwidth}
				X
			}
			\toprule
			Item & Choice & Consequence \\
			\midrule
			
			Gravity model
			&
			$f(R,T)=R+2\chi T$, with constant $\chi$
			&
			Since $f_R=1$, the vacuum exterior satisfies the Einstein vacuum equations.
			\\
			
			Metric and signature
			&
			$ds^2=e^{\nu}dt^2-e^{\lambda}dr^2-r^2d\Omega^2$
			&
			Determines the sign conventions entering the physical/effective-variable map and the surface stress.
			\\
			
			Matter source
			&
			Anisotropic
			$T^\mu{}_\nu=\mathrm{diag}(\rho,-p_r,-p_t,-p_t)$
			&
			Gives $T=\rho-p_r-2p_t$ and $\Delta=p_t-p_r$.
			\\
			
			Matter scalar
			&
			$L_m=-\avgP$, with
			$\avgP=q_\rho\rho+q_rp_r+q_tp_t$
			&
			The common-surface ratio depends on $(q_\rho,q_t)$; the model considered later uses $(0,1/3,2/3)$.
			\\
			
			Exterior
			&
			Schwarzschild spacetime with no exterior matter field
			&
			The surface compactness satisfies $C=M/R<1/2$.
			\\
			
			Smooth surface
			&
			Continuity of the first and second fundamental forms
			&
			Requires the effective normal traction to vanish and is therefore stronger than imposing only $p_r(R)=0$.
			\\
			
			Pseudo-spheroidal specialization
			&
			Eqs.~\eqref{eq:pseudometric} and \eqref{eq:prhat}
			&
			Leads to the exact compatibility relation $\chi_\partial(k,s;a,p_0)$.
			\\
			
			Tidal response
			&
			Static even-parity quadrupolar perturbation
			&
			Requires a prescription for the bulk anisotropic response and, when a shell is present, for the perturbed surface stresses.
			\\
			
			\bottomrule
		\end{tabularx}
	\end{table}
	\section{Smooth material and effective surfaces}
	\label{sec:smoothsurface}
	
	A material surface is defined by
	\begin{equation}
		p_r(R)=0.
		\label{eq:material_surface}
	\end{equation}
	A smooth matching to an exterior Schwarzschild spacetime, in the absence of a distributional surface stress, additionally requires the effective normal traction to vanish,
	$p_{r,\eff}(R)=0$
	together with continuity of the first and second fundamental forms. These two conditions are not generally equivalent in $f(R,T)$ gravity. It is therefore necessary to distinguish the material surface from the effective surface when discussing the stellar boundary.
	
	\begin{lemma}[Finite-density common-surface ratio]
		\label{lem:ratio}
		
		Let $\chi\neq0$, and suppose that the material and effective radial pressures vanish at the same radius $R$,
		\begin{equation}
			p_r(R)=p_{r,\eff}(R)=0.
			\label{eq:common_surface}
		\end{equation}
		For the general matter scalar
		\begin{equation}
			\avgP=q_\rho\rho+q_rp_r+q_tp_t,
			\qquad
			L_m=-\avgP,
			\label{eq:surface_matter_scalar}
		\end{equation}
		a nonzero surface density satisfies
		\begin{equation}
			\frac{p_t(R)}{\rho(R)}
			=
			\frac{p_{t,\eff}(R)}{\rho_\eff(R)}
			\equiv
			\kappa_s
			=
			\frac{1+2q_\rho}{2(1-q_t)},
			\label{eq:kappa_general}
		\end{equation}
		provided $q_t\neq1$ and the physical-to-effective map is nonsingular.
	\end{lemma}
	
	\begin{proof}
		
		At the material surface, $p_r(R)=0$. The effective radial pressure obtained from the field equations is
		\begin{equation}
			8\pi p_{r,\eff}
			=
			(8\pi+2\chi)p_r
			-\chi\left(T+2\avgP\right).
			\label{eq:pr_eff_general_surface}
		\end{equation}
		Therefore, at $r=R$,
		\begin{equation}
			8\pi p_{r,\eff}(R)
			=
			-\chi
			\left[
			T(R)+2\avgP(R)
			\right].
			\label{eq:pr_eff_at_surface}
		\end{equation}
		
		Using
		\begin{equation}
			T=\rho-p_r-2p_t,
			\label{eq:T_surface_definition}
		\end{equation}
		together with $p_r(R)=0$, we obtain
		\begin{equation}
			T(R)+2\avgP(R)
			=
			(1+2q_\rho)\rho(R)
			+
			2(q_t-1)p_t(R).
			\label{eq:TplusP_surface}
		\end{equation}
		
		The conditions $p_{r,\eff}(R)=0$ and $\chi\neq0$ therefore imply
		\begin{equation}
			(1+2q_\rho)\rho(R)
			+
			2(q_t-1)p_t(R)
			=
			0.
			\label{eq:surface_material_relation}
		\end{equation}
		For $q_t\neq1$ and $\rho(R)\neq0$, this gives
		\begin{equation}
			\frac{p_t(R)}{\rho(R)}
			=
			\frac{1+2q_\rho}{2(1-q_t)}
			\equiv
			\kappa_s.
			\label{eq:kappa_physical}
		\end{equation}
		
		Under the same condition,
		\begin{equation}
			T(R)+2\avgP(R)=0.
			\label{eq:surface_trace_condition}
		\end{equation}
		The field equations at the surface therefore reduce to
		\begin{equation}
			8\pi T^{\eff}_{\mu\nu}(R)
			=
			(8\pi+2\chi)T_{\mu\nu}(R).
			\label{eq:surface_effective_scaling}
		\end{equation}
		Consequently,
		\begin{equation}
			\rho_\eff(R)
			=
			\left(1+\frac{\chi}{4\pi}\right)\rho(R),
			\label{eq:surface_rho_scaling}
		\end{equation}
		and
		\begin{equation}
			p_{t,\eff}(R)
			=
			\left(1+\frac{\chi}{4\pi}\right)p_t(R).
			\label{eq:surface_pt_scaling}
		\end{equation}
		As long as this common factor is nonzero,
		\begin{equation}
			\frac{p_{t,\eff}(R)}{\rho_\eff(R)}
			=
			\frac{p_t(R)}{\rho(R)}
			=
			\kappa_s.
			\label{eq:surface_ratio_effective}
		\end{equation}
		This establishes the finite-density common-surface ratio.
	\end{proof}
	
	The surface ratio $\kappa_s$ is therefore fixed by the matter-Lagrangian prescription rather than by the interior geometry. Neither $\nu(r)$ nor $\lambda(r)$ enters the derivation. The coefficient $q_r$ also does not contribute because its term is proportional to $p_r(R)$, which vanishes at the material surface.
	
	For the commonly used average-pressure prescription,
	\begin{equation}
		L_m=-\frac{p_r+2p_t}{3},
		\label{eq:Lm_average_surface}
	\end{equation}
	one has
	\begin{equation}
		\kappa_s=\frac{3}{2}.
		\label{eq:kappa_average}
	\end{equation}
	For
	\begin{equation}
		L_m=-p_r,
		\label{eq:Lm_pr_surface}
	\end{equation}
	the corresponding ratio is
	\begin{equation}
		\kappa_s=\frac{1}{2}.
		\label{eq:kappa_pr}
	\end{equation}
	For
	\begin{equation}
		L_m=-p_t,
		\label{eq:Lm_pt_surface}
	\end{equation}
	one has $q_t=1$. In this case the denominator in the expression for $\kappa_s$ vanishes, and the common-surface condition instead reduces to
	\begin{equation}
		\rho(R)=0.
		\label{eq:rho_zero_pt_lagrangian}
	\end{equation}
	
	The common-surface ratio gives an algebraic condition at $r=R$. A further requirement follows from the behaviour of the effective radial pressure close to the surface. If
	\begin{equation}
		p_{r,\eff}(R)=0,
	\end{equation}
	and the effective radial pressure remains nonnegative immediately inside the star, then its first-order expansion is
	\begin{equation}
		p_{r,\eff}(r)
		=
		p'_{r,\eff}(R)(r-R)
		+
		\Order\!\left((r-R)^2\right).
		\label{eq:surface_pressure_expansion}
	\end{equation}
	For an interior point $r=R-\epsilon$, with $\epsilon>0$,
	\begin{equation}
		p_{r,\eff}(R-\epsilon)
		\simeq
		-\epsilon\,p'_{r,\eff}(R).
		\label{eq:surface_pressure_inside}
	\end{equation}
	A nonnegative pressure approaching zero from the stellar interior therefore requires, at first order,
	\begin{equation}
		p'_{r,\eff}(R)\leq0.
		\label{eq:surface_slope_condition}
	\end{equation}
	
	\begin{theorem}[Boundary-slope compactness bound]
		\label{thm:slope}
		
		Consider a static anisotropic configuration matched at $r=R$ to a Schwarzschild exterior, with compactness
		\begin{equation}
			C=\frac{M}{R}<\frac{1}{2}.
			\label{eq:compactness_definition_surface}
		\end{equation}
		Assume
		\begin{equation}
			p_{r,\eff}(R)=0,
			\qquad
			\rho_\eff(R)>0,
			\qquad
			p_{t,\eff}(R)=\kappa_s\rho_\eff(R).
			\label{eq:slope_theorem_assumptions}
		\end{equation}
		If $p_{r,\eff}$ remains nonnegative immediately inside the star and approaches zero with a nonpositive surface slope, then for $\kappa_s>0$,
		\begin{equation}
			\boxed{
				C\geq\frac{2\kappa_s}{1+4\kappa_s}
			}.
			\label{eq:Cbound}
		\end{equation}
		Equality corresponds to
		\begin{equation}
			p'_{r,\eff}(R)=0.
			\label{eq:marginal_surface_slope}
		\end{equation}
	\end{theorem}
	
	\begin{proof}
		
		At the smooth effective surface,
		\begin{equation}
			m(R)=M,
			\qquad
			p_{r,\eff}(R)=0.
			\label{eq:surface_mass_pressure}
		\end{equation}
		The radial derivative of the temporal metric potential at the surface is therefore
		\begin{equation}
			\frac{\nu'(R)}{2}
			=
			\frac{M}{R(R-2M)}.
			\label{eq:nu_surface}
		\end{equation}
		Using $C=M/R$, this becomes
		\begin{equation}
			\frac{R\nu'(R)}{2}
			=
			\frac{C}{1-2C}.
			\label{eq:nu_surface_compactness}
		\end{equation}
		
		At $r=R$, the effective equilibrium equation becomes
		\begin{equation}
			p'_{r,\eff}(R)
			=
			-\frac{\nu'(R)}{2}\rho_\eff(R)
			+
			\frac{2}{R}p_{t,\eff}(R).
			\label{eq:surface_tov}
		\end{equation}
		Multiplying by $R$ gives
		\begin{align}
			R\,p'_{r,\eff}(R)
			&=
			-\frac{C}{1-2C}\rho_\eff(R)
			+2p_{t,\eff}(R)
			\nonumber\\
			&=
			\rho_\eff(R)
			\left[
			2\kappa_s-\frac{C}{1-2C}
			\right].
			\label{eq:surface_slope}
		\end{align}
		
		Because $\rho_\eff(R)>0$, the condition $p'_{r,\eff}(R)\leq0$ requires
		\begin{equation}
			2\kappa_s
			\leq
			\frac{C}{1-2C}.
			\label{eq:slope_inequality}
		\end{equation}
		Since $C<1/2$, the denominator is positive. Hence,
		\begin{equation}
			2\kappa_s(1-2C)\leq C,
			\label{eq:slope_rearranged_1}
		\end{equation}
		which can be written as
		\begin{equation}
			2\kappa_s
			\leq
			C(1+4\kappa_s).
			\label{eq:slope_rearranged_2}
		\end{equation}
		
		Therefore, we recover the compactness bound
		stated in Eq.~\eqref{eq:Cbound}, completing
		the proof.
	\end{proof}
	
	It is useful to express the same result in dimensionless form. Define the normalized surface slope as
	\begin{equation}
		\mathcal{S}_s
		\equiv
		\frac{R\,p'_{r,\eff}(R)}
		{\rho_\eff(R)}.
		\label{eq:normalized_surface_slope}
	\end{equation}
	Then
	\begin{equation}
		\boxed{
			\mathcal{S}_s
			=
			2\kappa_s
			-
			\frac{C}{1-2C}
		}.
		\label{eq:surface_slope_dimensionless}
	\end{equation}
	The first-order requirement for a finite-density common smooth surface is
	\begin{equation}
		\mathcal{S}_s\leq0.
		\label{eq:Ss_condition}
	\end{equation}
	
	An equivalent boundary-compatibility quantity may be defined as
	\begin{equation}
		\mathcal{B}_s
		\equiv
		\frac{C}{1-2C}
		-
		2\kappa_s.
		\label{eq:boundary_diagnostic}
	\end{equation}
	The required first-order boundary condition is then
	\begin{equation}
		\boxed{
			\mathcal{B}_s\geq0
		}.
		\label{eq:boundary_diagnostic_condition}
	\end{equation}
	
	The compactness condition may also be written as an upper limit on the surface ratio,
	\begin{equation}
		\boxed{
			\kappa_s
			\leq
			\frac{C}{2(1-2C)}
		}.
		\label{eq:kappa_upper_bound}
	\end{equation}
	Using
	\begin{equation}
		\kappa_s
		=
		\frac{1+2q_\rho}{2(1-q_t)},
	\end{equation}
	the minimum compactness becomes
	\begin{equation}
		\boxed{
			C_{\min}
			=
			\frac{1+2q_\rho}
			{3+4q_\rho-q_t}
		}.
		\label{eq:Cmin_q}
	\end{equation}
	This result holds under the same assumptions and for $\kappa_s>0$. The threshold therefore depends on $q_\rho$ and $q_t$, while $q_r$ does not enter the boundary condition.
	
	\begin{corollary}[Average-pressure matter Lagrangian]
		\label{cor:three_sevenths}
		
		For
		\begin{equation}
			L_m=-\frac{p_r+2p_t}{3},
			\label{eq:Lm_average_corollary}
		\end{equation}
		the common-surface ratio is
		\begin{equation}
			\kappa_s=\frac{3}{2}.
		\end{equation}
		A common smooth finite-density surface with nonnegative effective radial pressure therefore requires
		\begin{equation}
			\boxed{
				C\geq\frac{3}{7}\simeq0.428571
			}.
			\label{eq:Cbound_average}
		\end{equation}
		For $L_m=-p_r$, the corresponding condition is
		\begin{equation}
			\boxed{
				C\geq\frac{1}{3}
			}.
			\label{eq:Cbound_pr}
		\end{equation}
	\end{corollary}
	
	For the average-pressure prescription, the normalized surface slope becomes
	\begin{equation}
		\mathcal{S}_s
		=
		3-\frac{C}{1-2C}
		=
		\frac{3-7C}{1-2C}.
		\label{eq:Ss_average}
	\end{equation}
	Since $1-2C>0$ for $C<1/2$, the sign of the surface slope is determined by $3-7C$. Hence,
	\begin{equation}
		C<\frac{3}{7}
		\quad\Longrightarrow\quad
		p'_{r,\eff}(R)>0.
		\label{eq:below_three_sevenths}
	\end{equation}
	This behaviour is incompatible, at first order, with a positive effective radial pressure approaching zero from inside the star.
	
	At
	\begin{equation}
		C=\frac{3}{7},
		\label{eq:C_marginal_average}
	\end{equation}
	the first derivative vanishes,
	\begin{equation}
		p'_{r,\eff}(R)=0.
		\label{eq:slope_zero_average}
	\end{equation}
	This represents a marginal first-order case. It does not by itself guarantee an acceptable pressure profile because the behaviour immediately below the surface is then controlled by higher-order derivatives.
	
	For $L_m=-p_r$, one has $\kappa_s=1/2$, and
	\begin{equation}
		\mathcal{S}_s
		=
		1-\frac{C}{1-2C}
		=
		\frac{1-3C}{1-2C}.
		\label{eq:Ss_pr}
	\end{equation}
	The corresponding necessary condition is therefore
	\begin{equation}
		C\geq\frac{1}{3}.
	\end{equation}
	
	\paragraph{Dominant-energy-condition implication.}
	
	For the average-pressure matter prescription, the
	finite-density common-surface condition requires
	\begin{equation}
		p_t(R)=\frac{3}{2}\rho(R).
	\end{equation}
	For a positive physical surface density,
	$\rho(R)>0$, this immediately gives
	\begin{equation}
		\rho(R)-|p_t(R)|
		=-\frac{1}{2}\rho(R)<0.
	\end{equation}
	Thus, the tangential dominant energy condition is
	violated at the common surface. Likewise, since
	$p_{t,\mathrm{eff}}(R)=3\rho_{\mathrm{eff}}(R)/2$,
	the effective tangential dominant energy condition
	is violated when $\rho_{\mathrm{eff}}(R)>0$.
	
	Consequently, satisfying the compactness requirement
	$C\geq 3/7$ does not establish that a finite-density
	common-surface configuration is physically admissible.
	The derived bound is a necessary local boundary
	compatibility condition under the stated assumptions,
	not a sufficient condition for satisfying the energy
	conditions or for constructing a globally acceptable
	stellar solution.
	
	The compactness condition derived above is a local boundary-compatibility condition and should not be interpreted as a general compactness bound for all anisotropic stars. In particular, the value $C\geq3/7$ does not imply that every physically acceptable configuration in the present theory must satisfy this inequality. Rather, below this value the assumptions of a finite positive effective surface density, coincident material and effective surfaces, the surface ratio $p_{t,\eff}(R)=3\rho_\eff(R)/2$, and a nonnegative effective radial pressure immediately inside a smoothly matched Schwarzschild surface cannot all be satisfied simultaneously.
	
	The restriction therefore applies to this particular boundary class and not to the pseudo-spheroidal geometry itself. No pseudo-spheroidal metric has been used in deriving the compactness condition. Configurations below the threshold may still be possible if the surface density vanishes, if a different matter-Lagrangian prescription is adopted, if the material and effective surfaces do not coincide, or if the interior is matched to the Schwarzschild exterior through a thin shell.
	
	For numerical and semi-analytic calculations, the boundary diagnostic
	\begin{equation}
		\mathcal{B}_s
		=
		\frac{C}{1-2C}
		-
		2\kappa_s
	\end{equation}
	provides a direct preliminary test. Three cases can be distinguished:
	\begin{align}
		\mathcal{B}_s>0
		&\quad\Longrightarrow\quad
		p'_{r,\eff}(R)<0,
		\\
		\mathcal{B}_s=0
		&\quad\Longrightarrow\quad
		p'_{r,\eff}(R)=0,
		\\
		\mathcal{B}_s<0
		&\quad\Longrightarrow\quad
		p'_{r,\eff}(R)>0.
	\end{align}
	
	The first case is compatible with an effective radial pressure that decreases to zero at the boundary. The second is marginal at first order and requires examination of higher-order terms. The third is incompatible with a nonnegative effective radial pressure immediately inside the assumed finite-density common smooth surface.

	\begin{figure}[t]
		\centering
		\includegraphics[width=0.72\textwidth]
		{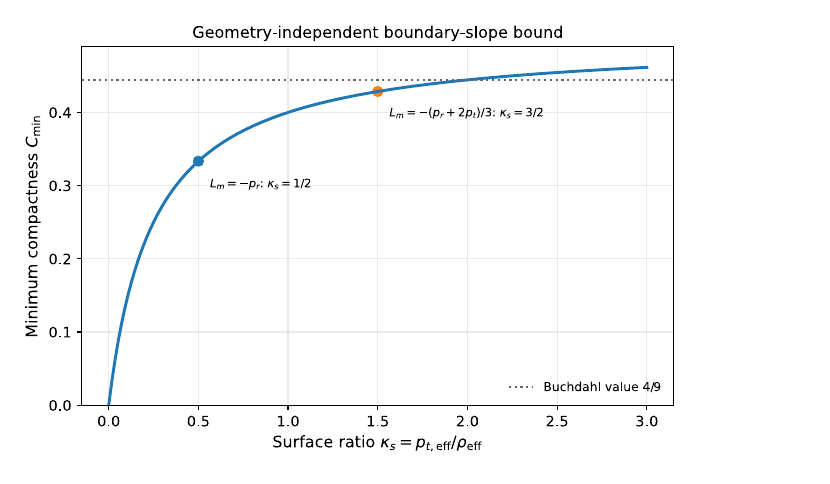}
		\caption{
			Geometry-independent lower compactness obtained from the boundary-slope condition. The curve
			$C_{\min}=2\kappa_s/(1+4\kappa_s)$
			separates finite-density common surfaces for which the effective radial pressure can approach zero with a nonpositive first-order slope from those for which this condition cannot be satisfied. For $\kappa_s>0$, $C_{\min}$ increases with $\kappa_s$ and approaches $1/2$ as $\kappa_s\rightarrow\infty$. The average-pressure prescription gives $\kappa_s=3/2$ and hence $C_{\min}=3/7$, whereas $L_m=-p_r$ gives $\kappa_s=1/2$ and $C_{\min}=1/3$.
		}
		\label{fig:slopebound}
	\end{figure}

	\section{Thin-shell completion}
	\label{sec:shell}
	
	The smooth-boundary condition discussed in the preceding section
	need not hold at the material surface. In particular, the surface
	may satisfy
	\begin{equation}
		p_r(R)=0,
		\label{eq:shell_material_surface}
	\end{equation}
	while the effective radial pressure is nonzero,
	\begin{equation}
		p_{r,\eff}(R)\neq0.
		\label{eq:shell_effective_nonzero}
	\end{equation}
	Continuity of the metric at $r=R$ then does not, in general,
	ensure continuity of the extrinsic curvature. Within the
	Einstein-like effective formulation adopted here, the remaining
	jump is represented by a distributional surface stress.
	
	We assume that the interior and exterior metric coefficients
	agree at the matching radius and write their common surface
	metric factor as
	\begin{equation}
		f_R=1-\frac{2M}{R}>0,
		\label{eq:shell_fR}
	\end{equation}
	while allowing the radial derivative of the temporal potential
	to be discontinuous,
	\begin{equation}
		\nu'_+\neq\nu'_-.
		\label{eq:shell_nu_jump}
	\end{equation}
	Here $f_R$ denotes the surface metric factor; it is distinct
	from the derivative $\partial f(R,T)/\partial R=1$ of the
	gravity Lagrangian.
	
	We use the bulk signature $(+,-,-,-)$ and the induced
	signature $(+,-,-)$ on the timelike hypersurface $\Sigma$.
	The spacelike unit normal is taken in the direction of
	increasing areal radius on both sides, so that
	$n^\mu n_\mu=-1$. With $e^\mu{}_a$ denoting the tangent
	basis to $\Sigma$, our extrinsic-curvature and jump
	conventions are
	\begin{equation}
		K_{ab}=e^\mu{}_a e^\nu{}_b\nabla_\mu n_\nu,
		\qquad
		[K^a{}_b]=K^a{}_{b,+}-K^a{}_{b,-}.
		\label{eq:shell_K_convention}
	\end{equation}
	The corresponding Israel junction relation is
	\begin{equation}
		8\pi S^a{}_b
		=[K^a{}_b]-\delta^a{}_b[K].
		\label{eq:shell_Israel_junction}
	\end{equation}
	In these conventions, the \emph{effective} mixed surface
	stress tensor takes the form
	\begin{equation}
		S^a{}_b
		=\operatorname{diag}
		\left(\sigma_s,-\mathcal P_s,-\mathcal P_s\right),
		\label{eq:Israel_surface_tensor}
	\end{equation}
	where $\sigma_s$ is the effective surface energy density and
	$\mathcal P_s$ is the effective surface pressure. A negative
	value of $\mathcal P_s$ represents surface tension.
	
	\begin{theorem}[Minimal surface layer for a continuous metric]
		\label{thm:shell}
		
		Under the above matching and sign conventions, continuity
		of $g_{tt}$ and $g_{rr}$ at $r=R$ implies
		\begin{equation}
			\sigma_s=0,
			\label{eq:shell_sigma_zero}
		\end{equation}
		whereas the effective surface pressure is
		\begin{equation}
			\boxed{
				\mathcal P_s
				=\frac{\sqrt{f_R}}{16\pi}
				\left(\nu'_+-\nu'_-\right)
				=-\frac{R\,p_{r,\eff}(R)}{2\sqrt{f_R}}
			}.
			\label{eq:shellpressure}
		\end{equation}
	\end{theorem}
	
	\begin{proof}
		For the static spherical hypersurface and the normal
		convention defined above, the nonzero mixed components
		of the extrinsic curvature are
		\begin{equation}
			K^t{}_{t,\pm}
			=\frac{\sqrt{f_R}}{2}\nu'_\pm,
			\qquad
			K^\theta{}_{\theta,\pm}
			=K^\phi{}_{\phi,\pm}
			=\frac{\sqrt{f_R}}{R}.
			\label{eq:shell_K_components}
		\end{equation}
		Since $f_R$ is continuous, the angular jumps vanish.
		Consequently, $[K]=[K^t{}_t]$, and the temporal
		component of Eq.~\eqref{eq:shell_Israel_junction} gives
		\begin{equation}
			\sigma_s=0.
			\label{eq:shell_sigma_proof}
		\end{equation}
		The angular component then gives
		\begin{equation}
			\mathcal P_s
			=\frac{\sqrt{f_R}}{16\pi}
			\left(\nu'_+-\nu'_-\right).
			\label{eq:shell_pressure_jump}
		\end{equation}
		
		For the Schwarzschild exterior,
		\begin{equation}
			\nu'_+
			=\frac{2M}{R(R-2M)}.
			\label{eq:shell_nu_plus}
		\end{equation}
		The interior Einstein-like radial field equation gives
		\begin{equation}
			\nu'_-
			=\frac{2\left[M+4\pi R^3p_{r,\eff}(R)\right]}
			{R(R-2M)}.
			\label{eq:shell_nu_minus}
		\end{equation}
		Hence the discontinuity in the temporal derivative is
		governed by the residual effective radial traction.
		Substitution in Eq.~\eqref{eq:shell_pressure_jump} yields
		\begin{equation}
			\mathcal P_s
			=-\frac{R\,p_{r,\eff}(R)}{2\sqrt{f_R}},
			\label{eq:shell_pressure_final}
		\end{equation}
		completing the proof.
	\end{proof}
	
	Equation~\eqref{eq:shell_pressure_final} relates the
	background effective shell stress directly to the
	interior effective radial traction. For
	\begin{equation}
		p_{r,\eff}(R)>0,
		\label{eq:shell_positive_traction}
	\end{equation}
	one obtains
	\begin{equation}
		\mathcal P_s<0,
		\label{eq:shell_tension_condition}
	\end{equation}
	so the matching requires an effective surface tension.
	Conversely, negative effective radial traction yields
	positive surface pressure. In the smooth-matching limit,
	\begin{equation}
		p_{r,\eff}(R)=0,
		\label{eq:shell_smooth_limit_pr}
	\end{equation}
	one recovers
	\begin{equation}
		\mathcal P_s=0.
		\label{eq:shell_smooth_limit_pressure}
	\end{equation}
	
	\paragraph{Energy conditions and interpretation of the surface layer.}
	
	For the continuous-metric shell, $\sigma_s=0$. When the
	residual effective radial traction is positive,
	Eq.~\eqref{eq:shellpressure} implies
	$\mathcal P_s<0$, and hence
	\begin{equation}
		\sigma_s+\mathcal P_s
		=\mathcal P_s<0.
		\label{eq:shell_NEC_failure}
	\end{equation}
	The effective shell therefore violates the surface null
	energy condition and, consequently, the weak energy
	condition. It also violates the surface dominant energy
	condition, since
	\begin{equation}
		\sigma_s=0<|\mathcal P_s|.
		\label{eq:shell_DEC_failure}
	\end{equation}
	The small numerical magnitude of a surface tension does
	not alter these sign inequalities.
	
	The tensor $S^a{}_b$ used here is the distributional
	\emph{Einstein-side effective source} defined by the
	extrinsic-curvature jump. In a matter--geometry-coupled
	theory it cannot, without an additional surface
	matter-variation prescription, be identified with the
	physical distributional matter stress tensor. The energy-
	condition statements above apply to this effective Israel
	layer. They do not, by themselves, establish the energy
	conditions of a separately specified physical matter shell.
	
	The effective surface layer is fixed by the chosen
	background matching; it is not an independently introduced
	bulk pressure contribution. However, its static stress does
	not specify its perturbative response. Under perturbations,
	the surface may acquire variations of its energy density and
	pressure, a displacement, and additional surface stresses.
	These require a consistent surface constitutive law and
	perturbed junction conditions. Thus the present geometric
	completion closes the \emph{background} matching problem
	within the adopted effective formulation, but does not by
	itself close the tidal perturbation problem.
	
	\section{Pseudo-spheroidal specialisation}
	\label{sec:pseudo}
	
	\subsection{Metric and prescribed effective pressure}
	
	We now apply the general results to a pseudo-spheroidal interior geometry described by the radial metric potential
	\begin{equation}
		e^{\lambda(r)}
		=
		\frac{1+k r^2/L^2}{1+r^2/L^2},
		\qquad
		k>1,
		\label{eq:pseudometric}
	\end{equation}
	where $k$ controls the spheroidal deformation and $L$ sets the radial length scale. It is convenient to introduce the dimensionless variables
	\begin{equation}
		x=\frac{r}{L},
		\qquad
		s=\frac{R}{L}.
		\label{eq:dimensionless_xs}
	\end{equation}
	
	For this geometry, the effective density takes the form
	\begin{equation}
		8\pi L^2\rho_\eff(x)
		=
		\widehat{\rho}(x)
		=
		\frac{(k-1)(3+kx^2)}
		{(1+kx^2)^2}.
		\label{eq:rhohat}
	\end{equation}
	
	The closure adopted in the present construction specifies the effective radial pressure as
	\begin{equation}
		8\pi L^2p_{r,\eff}(x)
		=
		\widehat p_r(x)
		=
		p_0
		\frac{(1-a x^2)(1+x^2)}
		{(1+kx^2)^2},
		\label{eq:prhat}
	\end{equation}
	where $a$ and $p_0$ are dimensionless parameters controlling the radial-pressure profile.
	
	Substituting this prescribed pressure into the effective radial field equation determines the derivative of the temporal metric potential,
	\begin{equation}
		\nu'(r)
		=
		\frac{p_0 r(L^2-ar^2)}
		{L^2(L^2+kr^2)}
		+
		\frac{(k-1)r}
		{L^2+r^2}.
		\label{eq:nup_pseudo}
	\end{equation}
	
	The effective tangential pressure is then obtained from the angular Einstein equation once $\nu(r)$ and $\lambda(r)$ are specified.
	
	Matching the interior radial metric component to the Schwarzschild exterior at $r=R$ gives the compactness relation
	\begin{equation}
		C\equiv\frac{M}{R}
		=
		\frac{(k-1)s^2}
		{2(1+ks^2)}.
		\label{eq:Cks}
	\end{equation}
	
	Thus, the compactness is determined by the spheroidicity parameter $k$ together with the scale ratio $s=R/L$. For fixed $s$, $C$ increases monotonically with $k$. This relation will be used below to express the surface compatibility condition directly in terms of the pseudo-spheroidal parameters.
	
	\subsection{Exact surface-selected coupling}
	
	The physical stellar boundary is defined by
	\begin{equation}
		p_r(R)=0.
		\label{eq:physical_surface_condition}
	\end{equation}
	Using the inverse relation between the physical and effective variables, this condition can be written as
	\begin{equation}
		3\chi\widehat\rho(s)
		+
		(11\chi+24\pi)\widehat p_r(s)
		-
		2\chi\widehat p_t(s)
		=
		0.
		\label{eq:surfaceeq}
	\end{equation}
	
	For the pseudo-spheroidal geometry and the prescribed effective radial-pressure profile introduced above, this surface condition determines the coupling parameter once the remaining model parameters are specified.
	
	\begin{theorem}[Surface-selected coupling]
		\label{thm:chi}
		
		For the pseudo-spheroidal closure defined by the radial metric and effective pressure profile, and away from singular branches, the material surface selects
		\begin{equation}
			\boxed{
				\chi_\partial(k,s;a,p_0)
				=
				\frac{
					48\pi p_0(1+s^2)^2(1-a s^2)(1+k s^2)
				}
				{\calD(k,s;a,p_0)}
			}.
			\label{eq:chi_general}
		\end{equation}
		Here, $\calD(k,s;a,p_0)$ is a quartic polynomial in $k$, given explicitly in Appendix~\ref{app:poly}.
		
		For the commonly used choice $s=1$, the expression reduces to
		\begin{equation}
			\chi_\partial(k;a,p_0)
			=
			\frac{
				192\pi p_0(1-a)(k+1)
			}
			{D_{a,p_0}(k)}.
			\label{eq:chi_s1}
		\end{equation}
		The denominator is
		\begin{align}
			D_{a,p_0}(k)
			=&\;
			k^4-10k^3
			+
			\left[4p_0(1-a)-40\right]k^2
			\nonumber\\
			&+
			\left[p_0(76a-92)+10\right]k
			+
			4p_0^2(a-1)^2
			\nonumber\\
			&+
			p_0(56a-72)
			+
			39.
			\label{eq:D_s1}
		\end{align}
		
		For the parameter values
		\begin{equation}
			a=0.8,
			\qquad
			p_0=1.5,
			\label{eq:special_ap0}
		\end{equation}
		the result becomes
		\begin{equation}
			\boxed{
				\chi_\partial(k)
				=
				\frac{
					1440\pi(k+1)
				}
				{
					25k^4-250k^3-970k^2-920k-36
				}
			}.
			\label{eq:chi_special}
		\end{equation}
		
	\end{theorem}
	
	\begin{proof}
		
		The result follows by substituting the effective density, the prescribed effective radial pressure, and the effective tangential pressure into the physical surface condition. The relevant effective density and radial-pressure profiles are
		\begin{equation}
			\widehat\rho(x)
			=
			\frac{(k-1)(3+kx^2)}
			{(1+kx^2)^2},
			\label{eq:rhohat_surface_recall}
		\end{equation}
		and
		\begin{equation}
			\widehat p_r(x)
			=
			p_0
			\frac{(1-a x^2)(1+x^2)}
			{(1+kx^2)^2}.
			\label{eq:prhat_surface_recall}
		\end{equation}
		The tangential effective pressure is obtained from the angular field equation using the pseudo-spheroidal metric and the corresponding expression for $\nu'(r)$.
		
		After the common denominators are removed, the resulting surface equation is linear in $\chi$. Solving this equation for $\chi$ gives the general surface-selected relation
		\begin{equation}
			\chi
			=
			\chi_\partial(k,s;a,p_0),
		\end{equation}
		with the explicit form given in Eq.~\eqref{eq:chi_general}. Collecting the powers of $k$ produces the quartic polynomial $\calD(k,s;a,p_0)$ listed in Appendix~\ref{app:poly}.
		
		Setting
		\begin{equation}
			s=1
		\end{equation}
		and simplifying gives Eq.~\eqref{eq:chi_s1}, together with the polynomial in Eq.~\eqref{eq:D_s1}. Finally, substituting
		\begin{equation}
			a=\frac{4}{5},
			\qquad
			p_0=\frac{3}{2},
		\end{equation}
		gives the explicit expression in Eq.~\eqref{eq:chi_special}.
		
	\end{proof}
	
	The important point is that, within this closure, $\chi$ is not independent of the stellar geometry once the material surface condition is imposed. Instead, the boundary condition defines a relation between $\chi$, $k$, $s$, $a$, and $p_0$.
	
	\begin{figure}[t]
		\centering
		\includegraphics[width=0.86\textwidth]
		{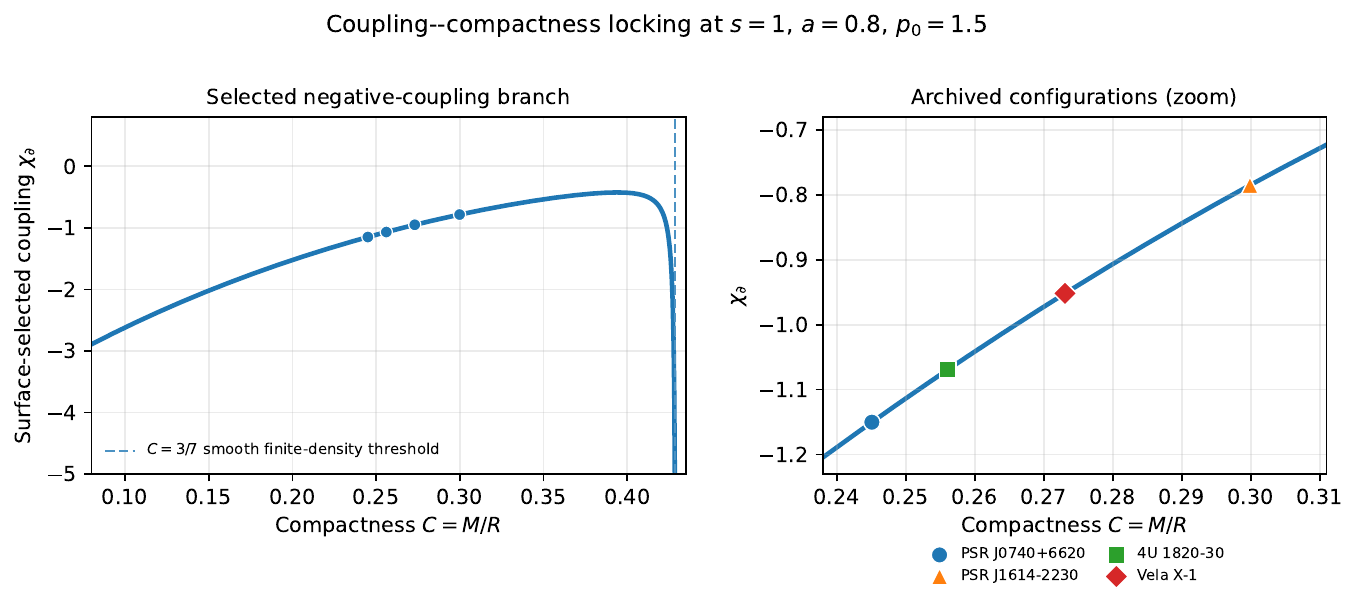}

		\caption{
			Selected negative-coupling branch of the
			surface-selected coupling relation for
			$s=1$, $a=0.8$, and $p_0=1.5$.
			The left panel shows the displayed
			negative-coupling branch and the
			finite-density smooth-surface threshold
			$C=3/7$. The right panel enlarges
			the interval containing the four
			archived stellar calibrations.
			The coupling has a pole at approximately
			$C=0.429384$, beyond which the
			algebraic relation continues on another
			branch not displayed here.
			The four points are boundary-selected
			calibrations, not independent
			determinations of a universal
			gravitational coupling.
		}

		\label{fig:locking}
	\end{figure}
	
	Figure~\ref{fig:locking} shows the surface-selected relation for
	\begin{equation}
		s=1,
		\qquad
		a=0.8,
		\qquad
		p_0=1.5.
	\end{equation}
	
	Figure~\ref{fig:locking} displays the selected
	negative-coupling branch of the surface-selected
	relation, together with the smooth finite-density
	threshold at $C=3/7$. The four archived stellar
	configurations lie on this branch and represent
	compactness-dependent boundary calibrations
	rather than independent determinations of a
	universal gravitational coupling.
	
	The algebraic relation has a pole at approximately
	$C=0.429384$, beyond which it continues on
	another branch. This continuation is not displayed
	in Fig.~\ref{fig:locking}. The existence of
	additional algebraic branches does not, by itself,
	establish their physical admissibility.

	\begin{figure}[t]
		\centering
		\includegraphics[width=0.78\textwidth]
		{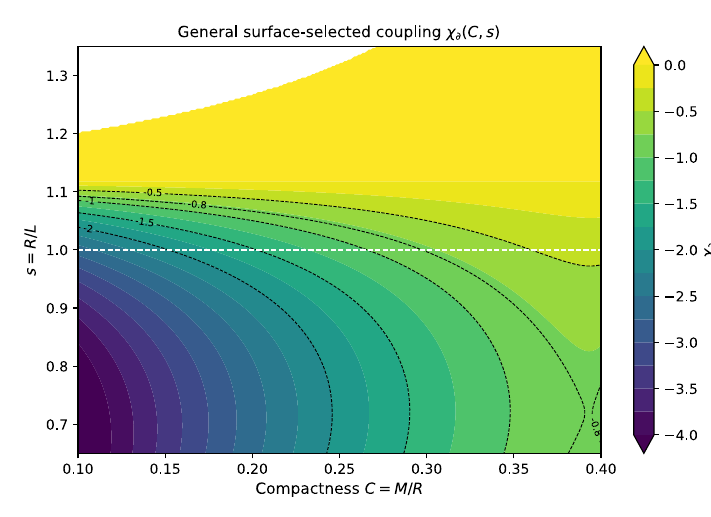}
		
		\caption{
			General surface-selected coupling
			$\chi_\partial(C,s)$ for $a=0.8$
			and $p_0=1.5$. The displayed colour
			scale is restricted to
			$-4\leq\chi_\partial\leq0$.
			Values outside this range are not
			quantitatively resolved by the
			colour mapping. Saturated or unfilled
			regions should not be interpreted as
			physically excluded configurations.
			Singular points where
			$\mathcal D(k,s;a,p_0)=0$ are not
			regular values of the explicitly
			solved coupling relation.
			The contours illustrate algebraic
			boundary compatibility rather than
			physical admissibility.
		}
		\label{fig:chimap}
	\end{figure}
	
	Figure~\ref{fig:chimap} extends the same analysis by allowing
	\begin{equation}
		s=\frac{R}{L}
	\end{equation}
	to vary. Releasing the condition $s=1$ introduces an additional dimensionless degree of freedom, so the compatibility relation becomes a surface $\chi_\partial(C,s)$. Even in this more general case, a fixed value of $\chi$ selects level sets on that surface rather than an unrestricted family of models. The physical admissibility conditions must therefore be examined along the corresponding level sets.

	\subsection{Why the coupling is consumed: a degree-of-freedom count}
	
	The origin of the coupling--compactness relation can be understood without using the explicit quartic polynomial. Once the condition
	\begin{equation}
		s=1
	\end{equation}
	is imposed, matching $g_{rr}$ to the Schwarzschild exterior assigns a unique value of the spheroidicity parameter $k$ to each chosen compactness. If $a$ and $p_0$ are also fixed, the dimensionless effective density and pressure profiles are completely determined. Matching the temporal metric component then fixes the remaining multiplicative constant in $e^{\nu}$.
	
	At this stage, the material surface condition contains only one undetermined quantity, namely the coupling $\chi$. Because the transformation between the physical and effective variables is algebraic, the surface equation is linear in $\chi$. Solving this equation therefore does not provide an independent determination of the coupling. Instead, it gives the particular value of $\chi$ required by the chosen compactness and the adopted closure. The exact relation obtained in Theorem~\ref{thm:chi} is therefore the analytic form of this parameter counting, rather than a coincidence associated with a small number of numerical configurations.
	
	Allowing
	\begin{equation}
		s=\frac{R}{L}
	\end{equation}
	to vary restores one dimensionless structural degree of freedom. For a fixed value of $\chi$, the general surface relation then defines curves in the $(k,s)$ plane instead of isolated solutions. These curves represent compatibility level sets, but they are not automatically physically acceptable stellar families. Each point must still satisfy the required conditions on positivity, causality, regularity, nonsingular denominators, the adopted surface class, and any additional constitutive assumptions.
	
	Similarly, allowing $a$ or $p_0$ to vary separately for each stellar configuration increases the available fitting freedom. In that case, however, these parameters also become configuration-dependent calibration parameters and can no longer be interpreted directly as fixed properties of a single matter model.
	
	\subsection{Fixed universal coupling and compactness classes}
	
	The effect of fixing the gravitational coupling can be understood directly from the surface relation. Once $\chi$, $s$, $a$, and $p_0$ are specified, the material boundary condition becomes an algebraic equation for the spheroidicity parameter $k$.
	
	\begin{theorem}[Finite compactness classes at fixed coupling]
		\label{thm:finiteclasses}
		
		Let $(\chi,s,a,p_0)$ be fixed, with $\chi\neq0$, and assume that the denominators of the physical--effective map and of the surface-selected coupling remain nonzero. Then the surface condition is a polynomial equation of degree at most four in $k$. It follows that the closure can admit at most four real values of the spheroidicity parameter. Since, for fixed $s>0$, the compactness is a one-to-one function of $k$, the same model can contain at most four corresponding compactness classes.
		
	\end{theorem}
	
	\begin{proof}
		
		Starting from the exact surface-selected coupling and cross-multiplying gives
		\begin{equation}
			\chi\calD(k,s;a,p_0)
			-
			48\pi p_0(1+s^2)^2(1-a s^2)(1+k s^2)
			=
			0.
			\label{eq:fixedchi_polynomial}
		\end{equation}
		
		The first term contains a quartic polynomial in $k$, whereas the second term is only linear in $k$. The resulting equation is therefore of degree no higher than four and can have at most four real roots for $k$.
		
		The relation between compactness and $k$ is
		\begin{equation}
			C
			=
			\frac{(k-1)s^2}
			{2(1+ks^2)}.
			\label{eq:Cks_fixedchi}
		\end{equation}
		For fixed $s$, its derivative is
		\begin{equation}
			\frac{\partial C}{\partial k}
			=
			\frac{s^2(1+s^2)}
			{2(1+ks^2)^2}
			>0.
			\label{eq:dCdk_fixedchi}
		\end{equation}
		Hence, $C$ increases monotonically with $k$, so each real value of $k$ corresponds to one compactness value. The maximum number of compactness classes is therefore also four.
		
	\end{proof}
	
	This result explains why fixing one universal value of $\chi$ does not generally produce an arbitrary collection of observed compactnesses within the same closure. If a numerical scan at fixed $\chi$, $s$, $a$, and $p_0$ fails to recover several independently chosen stellar configurations, this need not indicate a numerical problem. It may instead follow directly from the algebraic structure of the surface condition.
	
	The representative solutions listed in Table~\ref{tab:fixedchi} illustrate this point for $s=1$, $a=0.8$, and $p_0=1.5$. For each fixed value of $\chi$, the surface equation gives isolated real values of $k$, and hence isolated compactnesses, rather than a continuous sequence of compactness values at fixed dimensionless parameters. In the examples shown, two real branches are present: one at lower compactness and another at higher compactness.
	
	This distinction is important when interpreting the model as a stellar family. A fixed value of the gravitational coupling does not by itself generate a continuous mass--radius sequence. Instead, within the present closure and for fixed dimensionless parameters, the boundary condition selects a finite set of compatible compactness classes.
	
	\begin{table}[t]
		\caption{Representative real roots of the fixed-$\chi$ surface equation for $s=1$, $a=0.8$, and $p_0=1.5$. For each fixed coupling, the surface condition selects isolated compactness classes rather than a continuous sequence at fixed dimensionless parameters.}
		\label{tab:fixedchi}
		\centering
		\begin{tabular}{
				S[table-format=-1.2]
				S[table-format=2.6]
				S[table-format=1.6]
				@{\qquad}
				S[table-format=2.6]
				S[table-format=1.6]
			}
			\toprule
			{$\chi$} & {$k_1$} & {$C_1$} & {$k_2$} & {$C_2$} \\
			\midrule
			-2.00 & 1.873356 & 0.151975 & 12.682045 & 0.426912 \\
			-1.50 & 2.362478 & 0.202600 & 12.505588 & 0.425957 \\
			-1.15 & 2.923167 & 0.245104 & 12.276370 & 0.424678 \\
			-1.00 & 3.272411 & 0.265940 & 12.119195 & 0.423776 \\
			-0.80 & 3.934084 & 0.297328 & 11.791256 & 0.421822 \\
			-0.50 & 6.141167 & 0.359967 & 10.406495 & 0.412331 \\
			\bottomrule
		\end{tabular}
	\end{table}
	
	\subsection{Scale homology}
	
	The role of the dimensional scale $L$ can be separated from the dimensionless structure of the model. If the dimensionless parameters
	\begin{equation}
		(k,s,a,p_0,\chi)
	\end{equation}
	are held fixed and only $L$ is varied, then the stellar radius and mass scale as
	\begin{equation}
		R=sL,
		\qquad
		M=C(k,s)R.
		\label{eq:homology_MR_scaling}
	\end{equation}
	All dimensionless metric functions and matter profiles remain unchanged.
	
\begin{figure}[t]
		\centering
		\includegraphics[width=0.72\textwidth]
		{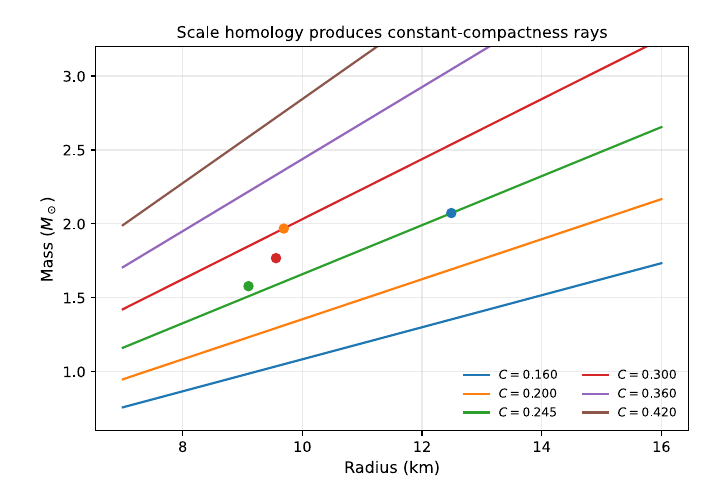}
		\caption{
			Scale-only variation at fixed dimensionless parameters produces rays $M=CR$ in the mass--radius plane. Points on a given ray have the same compactness and identical normalized profiles. They therefore form a homologous family rather than an equation-of-state sequence, unless a common dimensional constitutive relation is independently established.
		}
		\label{fig:homology}
	\end{figure}

	In Fig.~\ref{fig:homology}, the mass is expressed
	in solar masses and the radius in kilometres.
	Using $GM_\odot/c^2 \simeq 1.476625\,\mathrm{km}$,
	the plotted relation is
	\[
	\frac{M}{M_\odot}
	=
	\frac{C\,(R/\mathrm{km})}{1.476625}.
	\]

	\begin{theorem}[\textbf{Homology of scale-only families}]
		\label{thm:homology}
		
		For fixed $(k,s,a,p_0,\chi)$, varying only the scale $L$ generates a family of configurations with the same compactness and identical normalized profiles. The resulting configurations therefore form a homologous family. Without an additional dimensional constitutive relation, this family should not be interpreted as a stellar sequence obtained by varying the central density while keeping one dimensional equation of state fixed.
		
	\end{theorem}
	
	\begin{proof}
		
		The pseudo-spheroidal metric, effective density, and prescribed effective radial pressure depend on the radial coordinate only through the dimensionless variable
		\begin{equation}
			x=\frac{r}{L}.
			\label{eq:homology_dimensionless_radius}
		\end{equation}
		The dimensional density and pressure therefore scale as
		\begin{equation}
			\rho_{\eff}\propto L^{-2},
			\qquad
			p_{r,\eff}\propto L^{-2}.
			\label{eq:homology_density_pressure_scaling}
		\end{equation}
		
		At the same time, the compactness relation
		\begin{equation}
			C=C(k,s)
			\label{eq:homology_compactness}
		\end{equation}
		contains no explicit dependence on $L$. Consequently, increasing or decreasing $L$ rescales both $R$ and $M$ linearly, while leaving the ratio $M/R$ and all normalized profiles unchanged.
		
	\end{proof}
	
	This behaviour is illustrated by the mass--radius rays in Fig.~\ref{fig:homology}. Every point on a given ray satisfies
	\begin{equation}
		M=CR
		\label{eq:homology_ray}
	\end{equation}
	with the same compactness. Such points therefore differ only by their overall length scale and belong to the same homologous class. They should not be interpreted as a conventional equation-of-state sequence unless a common dimensional constitutive law is established independently.
	
	\subsection{Common physical/effective surface within the prescribed profile}
	
	The prescribed effective radial-pressure profile vanishes at the boundary when
	\begin{equation}
		a s^2=1.
		\label{eq:common_surface_as}
	\end{equation}
	
	If the physical radial pressure is required to vanish at the same radius,
	\begin{equation}
		p_r(R)=0,
		\label{eq:common_surface_pr_zero}
	\end{equation}
	then the physical and effective surfaces coincide. Imposing these two conditions simultaneously gives an exact relation for the profile parameter $p_0$.
	
	\begin{corollary}[Common-surface profile parameter]
		\label{cor:common}
		
		For
		\begin{equation}
			a s^2=1,
			\label{eq:common_surface_condition}
		\end{equation}
		a nontrivial common physical/effective surface requires
		\begin{equation}
			\boxed{
				p_0
				=
				\frac{
					(k-1)(ks^2+3)(ks^2-7s^2-6)
				}{
					4(1+s^2)^2
				}
			}.
			\label{eq:p0_common}
		\end{equation}
		
		A positive value of $p_0$ is possible only when
		\begin{equation}
			C>\frac{3}{7}.
			\label{eq:common_surface_compactness_bound}
		\end{equation}
		
		The threshold $C=3/7$ agrees with that obtained independently from the general boundary-slope condition, but the strictness of the two conditions is different.
		
	\end{corollary}
	
	The agreement provides an internal consistency check of the construction. The present condition follows directly from the prescribed effective-pressure profile together with the requirement that the material and effective surfaces coincide.
	
	The general boundary-slope condition gives
	\begin{equation}
		C\geq\frac{3}{7}.
		\label{eq:common_surface_general_bound}
	\end{equation}
	For the prescribed common-surface profile, however, requiring $p_0>0$ strengthens this condition to
	\begin{equation}
		C>\frac{3}{7},
	\end{equation}
	while equality,
	\begin{equation}
		C=\frac{3}{7},
	\end{equation}
	corresponds to
	\begin{equation}
		p_0=0.
	\end{equation}
	
	Thus, the general boundary-slope result and the positive-$p_0$ condition share the same compactness threshold but differ in whether equality is allowed. This compactness condition remains a necessary local boundary-compatibility requirement under the stated assumptions and should not be interpreted as a universal compactness bound.
	
\begin{figure}[t]
		\centering
		\includegraphics[width=0.73\textwidth]
		{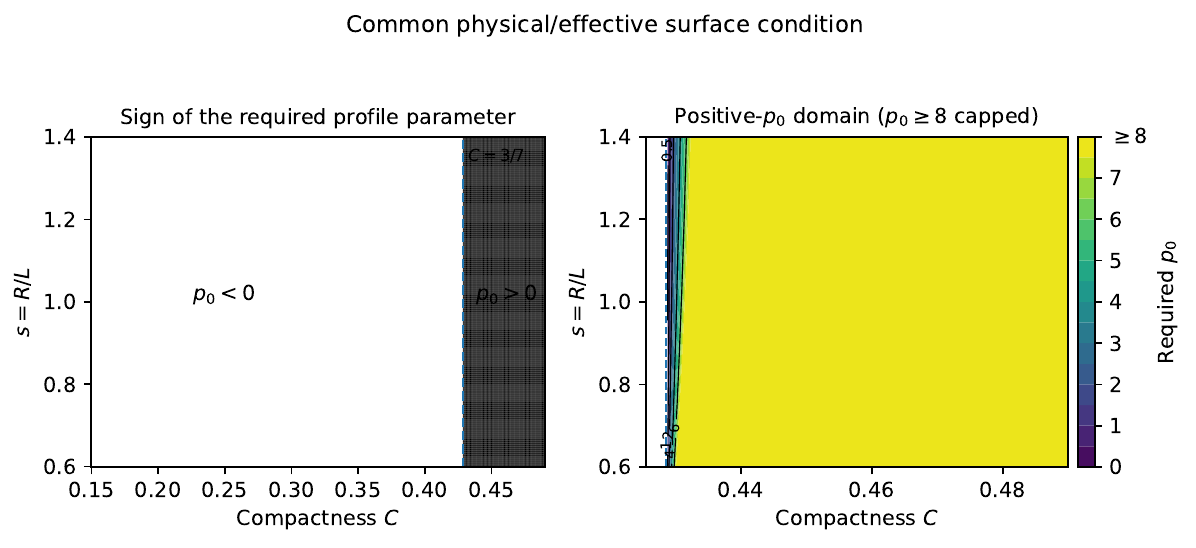}
		
		\caption{
			Algebraic common physical/effective surface condition determined by
			Eq.~\eqref{eq:p0_common}.
			The left panel distinguishes the regions with $p_0<0$ and $p_0>0$.
			The boundary of the positive-$p_0$ region occurs at $C=3/7$, where
			$p_0=0$, while $p_0>0$ requires $C>3/7$.
			The right-panel colour scale is capped at $p_0=8$, so its highest
			colour represents all $p_0\geq8$.
			The figure identifies an algebraic positive-$p_0$ domain and does not
			establish satisfaction of the dominant energy condition.
		}
		\label{fig:common}
	\end{figure}

	Figure~\ref{fig:common} illustrates the corresponding parameter domain. The boundary of the positive-$p_0$ region occurs at
	\begin{equation}
		C=\frac{3}{7},
	\end{equation}
	where $p_0=0$, while strictly positive values of $p_0$ occur for
	\begin{equation}
		C>\frac{3}{7}.
	\end{equation}
	
	\section{Four illustrative stellar calibrations}
	\label{sec:fourstars}

	The four examples below reproduce the input table from the earlier pseudo-spheroidal calculation. That table records the radial metric parameter and the reported coupling for each object. A second table from the same calculation records the adopted radius, compactness and nominal interior tidal input. We use these values to test the algebraic reconstruction in Table~\ref{tab:fourstars}. The original observational data releases and uncertainties for the adopted masses and radii have not been identified.

	Within the archived prescription, the radial length scale
	was set to
	\begin{equation}
		L=R,
		\label{eq:fourstars_LR}
	\end{equation}
	and the profile parameters were fixed at
	\begin{equation}
		(a,p_0)=(0.8,1.5).
		\label{eq:fourstars_ap0}
	\end{equation}
	
	Continuity of the radial metric coefficient with the
	Schwarzschild exterior then gives
	\begin{equation}
		k=\frac{1+2C}{1-2C},
		\qquad
		C=\frac{M}{R},
		\label{eq:fourstars_kC}
	\end{equation}
	where $M$ is expressed in geometrized length units.
	Equivalently,
	\begin{equation}
		C=\frac{k-1}{2(k+1)}.
	\end{equation}
	
	The archived $k$ values are quoted to six decimal places,
	whereas the compactnesses used in the shell and tidal
	illustrations are quoted to four decimal places.
	The small differences between $C$ reconstructed from $k$
	and the displayed compactness are therefore attributable
	to rounding, not to distinct stellar solutions.
	
	For PSR J0740+6620, PSR J1614--2230, 4U 1820--30,
	and Vela X--1, respectively, the archived radii are
	\begin{equation}
		R=(12.49,\,9.69,\,9.10,\,9.56)\,\mathrm{km},
	\end{equation}
	and the archived compactnesses, rounded to four decimal
	places, are
	\begin{equation}
		C=(0.2451,\,0.2999,\,0.2561,\,0.2731).
	\end{equation}
	
	Their corresponding masses are implied by $M=CR$;
	they are not independently inferred from an observational
	likelihood in the present analysis.
	
	The nominal interior-side tidal inputs adopted for the
	conditional sensitivity calculations are
	\begin{equation}
		y_-=(1.7242,\,1.3008,\,1.6801,\,1.5817)
	\end{equation}
	in the same object order.
	These values correspond, respectively, to
	PSR J0740+6620, PSR J1614--2230,
	4U 1820--30, and Vela X--1.
	They are adopted illustrative inputs rather than outputs
	of a documented integration of the bulk anisotropic
	tidal perturbation equations.
	They are used in Section~6 solely to evaluate the
	conditional sensitivity of the exterior Love-number
	formula to assumed surface jumps.
	
	After fixing the geometric and profile parameters,
	the material boundary condition
	\begin{equation}
		p_r(R)=0
		\label{eq:fourstars_material_surface}
	\end{equation}
	selects the remaining coupling $\chi$ through the
	algebraic surface relation $\chi_\partial(k)$.
	
\begin{figure}[t]
		\centering
		\includegraphics[width=0.92\textwidth]
		{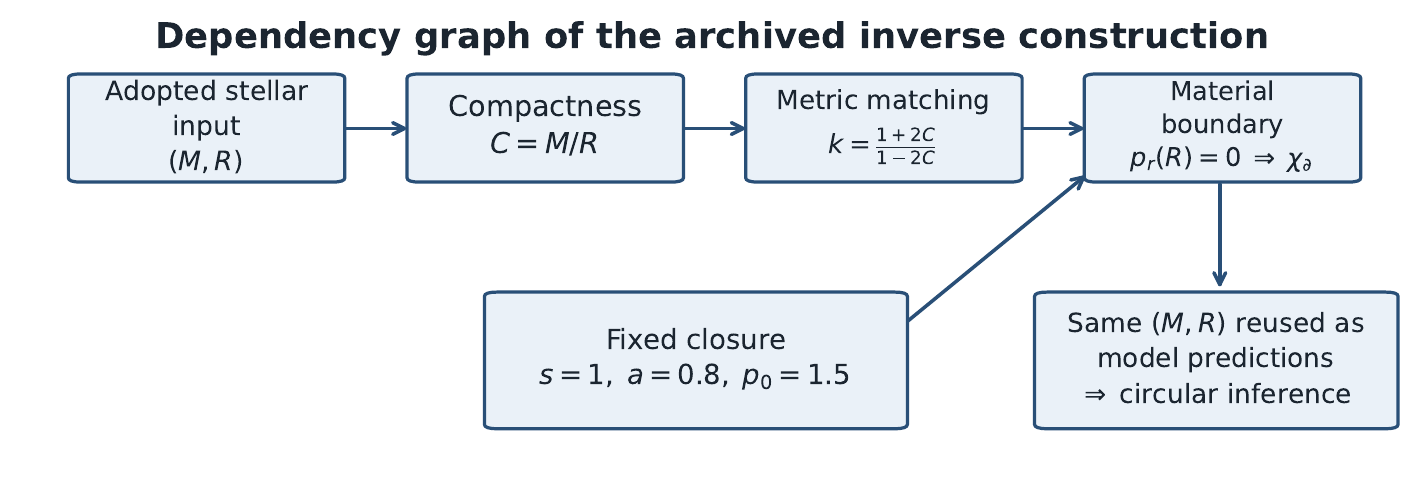}
		\caption{
			Dependency structure of the archived inverse
			construction. The adopted stellar mass and
			radius determine the compactness and the
			radial metric parameter. The fixed closure
			and material boundary condition then select
			the coupling. Reusing the input mass--radius
			values as independent model predictions
			would be circular.
		}
		\label{fig:workflow}
	\end{figure}

	This dependence is illustrated in Fig.~\ref{fig:workflow}.
	Since $k$ is already selected by the adopted compactness,
	the resulting $\chi$ values are
	\emph{boundary-selected calibrations}, not four independent
	measurements of a universal gravitational coupling.
	Nor may the same adopted $(M,R)$ values be counted again
	as predictions of the fitted model.
	
	The exact surface-selected relation reproduces the
	reported couplings at the precision of the archived
	$k$ and $\chi$ entries.
	Table~\ref{tab:fourstars} shows this arithmetic
	reproduction; the small residuals test the implementation
	of the algebraic formula, not the observational provenance
	of the adopted stellar parameters.
	
	\begin{table}[t]
		\caption{
			Reproduction of the four archived couplings
			from the exact surface-selected relation.
			The final column gives
			$\chi_\partial-\chi_{\rm tab}$.
			Residuals of order $10^{-7}$ or smaller are
			consistent with the displayed input precision;
			they do not represent uncertainties on
			the coupling.
		}
		\label{tab:fourstars}
		\centering
		{\small
			\setlength{\tabcolsep}{3pt}
			\begin{tabular}{
					l
					S[table-format=1.6]
					S[table-format=-1.6]
					S[table-format=-1.9]
					S[table-format=-1.2e-1]
				}
				\toprule
				Object
				& {$k$}
				& {$\chi_{\rm tab}$}
				& {$\chi_\partial$}
				& {$\chi_\partial-\chi_{\rm tab}$} \\
				\midrule
				
				PSR J0740+6620
				& 2.922793
				& -1.150183
				& -1.150183167
				& -1.67e-7 \\
				
				PSR J1614--2230
				& 3.997000
				& -0.785113
				& -0.785113004
				& -3.53e-9 \\
				
				4U 1820--30
				& 3.100000
				& -1.069130
				& -1.069130203
				& -2.03e-7 \\
				
				Vela X--1
				& 3.407000
				& -0.951717
				& -0.951717255
				& -2.55e-7 \\
				
				\bottomrule
			\end{tabular}
		}
	\end{table}
	
	The compactnesses of these four configurations occupy
	the interval
	\begin{equation}
		0.245\lesssim C\lesssim0.300,
		\label{eq:fourstars_compactness_range}
	\end{equation}
	which lies entirely below the necessary finite-density,
	common-surface threshold for the adopted average-pressure
	matter prescription,
	\begin{equation}
		C=\frac{3}{7}.
		\label{eq:fourstars_threshold}
	\end{equation}
	
	Consequently, under the regularity and nonsingularity
	assumptions used in the common-surface derivation,
	none of the four archived configurations satisfies the
	conditions for a smooth, finite-density surface at which
	both $p_r(R)$ and $p_{r,\eff}(R)$ vanish.
	
	This is a boundary-compatibility statement; the inequality
	by itself does not certify or exclude other possible
	material prescriptions or global stellar configurations.
	
	For the adopted effective radial-pressure profile at
	$L=R$ and $(a,p_0)=(0.8,1.5)$, one obtains directly
	\begin{equation}
		8\pi R^2 p_{r,\eff}(R)
		=
		\frac{2p_0(1-a)}{(1+k)^2}
		=
		\frac{0.6}{(1+k)^2}>0.
		\label{eq:fourstars_effective_traction_formula}
	\end{equation}
	
	Thus the nonzero effective radial traction is explicitly
	present in all four cases:
	\begin{equation}
		p_{r,\eff}(R)>0.
		\label{eq:fourstars_positive_preff}
	\end{equation}
	
	Within the effective Einstein-side junction construction,
	continuous matching of the induced metric to the
	Schwarzschild exterior then requires a surface layer.
	Its pressure is
	\begin{equation}
		\mathcal P_s
		=
		-\frac{R\,p_{r,\eff}(R)}
		{2\sqrt{f_R}},
		\qquad
		f_R=1-2C=\frac{2}{1+k}.
		\label{eq:fourstars_shell_pressure}
	\end{equation}
	
	Here $f_R$ denotes only the
	\emph{surface value of the Schwarzschild metric factor},
	not the derivative $\partial f/\partial R$ of the
	gravitational action.
	
	The sign is negative because $p_{r,\eff}(R)>0$,
	so $\mathcal P_s$ represents effective surface tension
	under the adopted junction convention.
	
	The corresponding dimensionless pressure is
	\begin{equation}
		R\mathcal P_s
		=
		-\frac{0.6}
		{16\pi(1+k)^2\sqrt{2/(1+k)}}.
		\label{eq:fourstars_shell_scale}
	\end{equation}
	
	Since $p_{r,\eff}(R)>0$ for all four archived
	configurations, the corresponding signed dimensionless
	surface pressure satisfies
	\begin{equation}
		R\mathcal P_s<0.
		\label{eq:shell_csv_sign}
	\end{equation}
	
\begin{figure}[t]
		\centering
		\includegraphics[width=0.75\textwidth]
		{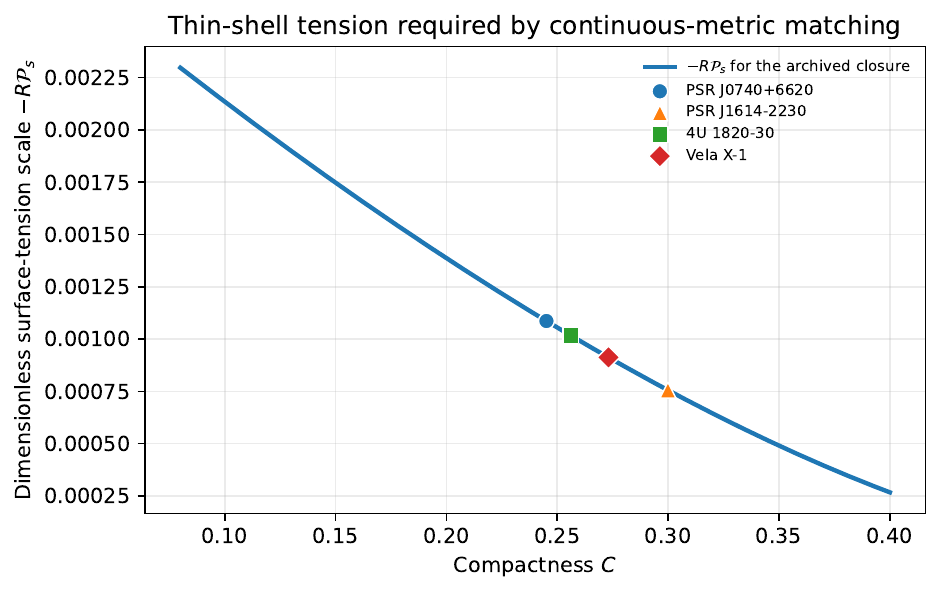}
		\caption{
			Magnitude of the negative background effective
			shell pressure, $-R\mathcal P_s$, plotted
			against the archived compactness.
			The positive ordinate denotes tension,
			not positive surface pressure.
			The curve and points follow the archived
			background formula; no perturbative surface
			constitutive law is supplied by this
			calculation.
		}
		\label{fig:shell}
	\end{figure}

	Table~\ref{tab:surfaceaudit} reports this negative
	signed pressure, whereas Fig.~\ref{fig:shell} displays
	its positive tension magnitude $-R\mathcal P_s$.
	The difference is entirely due to the sign convention
	and does not represent a discrepancy in the numerical
	calculations.
	
	\begin{table}[t]
		\caption{
			Background effective surface audit for the
			four archived calibrations.
			The displayed compactness $C$ is rounded,
			while $f_R$ and the surface quantities are
			evaluated from the archived $k$ and $R$.
			The pressure $p_{r,\eff}(R)$ is expressed
			in geometrized units, and negative
			$R\mathcal P_s$ denotes effective surface
			tension. The physical constitutive origin
			of this effective surface layer is not
			specified by the background calculation.
		}
		\label{tab:surfaceaudit}
		\centering
		\begin{tabular}{
				l
				S[table-format=1.4]
				S[table-format=1.6]
				S[table-format=1.3e-2]
				S[table-format=-1.3e-2]
			}
			\toprule
			Object
			& {$C$}
			& {$f_R$}
			& {$p_{r,\eff}(R)\,[\mathrm{km}^{-2}]$}
			& {$R\mathcal P_s$} \\
			\midrule
			
			PSR J0740+6620
			& 0.2451
			& 0.509841
			& 9.945e-6
			& -1.086e-3 \\
			
			PSR J1614--2230
			& 0.2999
			& 0.400240
			& 1.018e-5
			& -0.756e-3 \\
			
			4U 1820--30
			& 0.2561
			& 0.487805
			& 1.715e-5
			& -1.017e-3 \\
			
			Vela X--1
			& 0.2731
			& 0.453823
			& 1.345e-5
			& -0.912e-3 \\
			
			\bottomrule
		\end{tabular}
	\end{table}
	
	The layer is required at the background level because
	the material boundary condition $p_r(R)=0$ does not
	coincide with vanishing effective radial traction.
	Although its dimensionless pressure magnitude is small,
	the layer cannot simply be omitted from the junction
	problem.
	
	Finally, a static effective surface pressure does not
	determine the shell's response to a quadrupolar
	perturbation.
	A specified bulk tidal closure, a perturbed junction
	prescription, and an appropriate surface-response
	relation are required to obtain a physical exterior
	Love number.
	
	The archived $y_-$ inputs therefore remain conditional
	illustrative inputs, as discussed in the following
	tidal section.
	
	\FloatBarrier

	\section{Tidal closure and surface nonuniqueness}
	\label{sec:tides}
	
	\subsection{Bulk perturbation closure}
	
	For a static even-parity quadrupolar perturbation in
	Regge--Wheeler gauge, we write schematically
	\begin{equation}
		h_{\mu\nu}\dd x^\mu\dd x^\nu
		=
		e^{\nu}H(r)Y_{2m}\dd t^2
		+
		e^{\lambda}H(r)Y_{2m}\dd r^2
		+
		r^2K(r)Y_{2m}\dd\Omega^2.
	\end{equation}
	The perturbed field equations follow directly from
	Eq.~\eqref{eq:field}:
	\begin{equation}
		\delta G_{\mu\nu}
		=
		(8\pi+2\chi)\delta T_{\mu\nu}
		+
		\chi(\delta T+2\delta\avgP)g^{(0)}_{\mu\nu}
		+
		\chi(T+2\avgP)h_{\mu\nu}.
		\label{eq:pertfield}
	\end{equation}
	
	The equilibrium background does not by itself determine
	$\delta p_t$. A local linear constitutive closure may be written
	in the general form
	\begin{equation}
		\delta p_t
		=
		A(r)\,\delta p_r
		+
		B(r)\,H,
		\label{eq:closure}
	\end{equation}
	where $A(r)$ represents the material response to
	$\delta p_r$ at fixed geometric state, while $B(r)$ represents
	explicit metric or quasi-local dependence.
	For a quasi-local anisotropy
	$\Del=\Del(\rho,p_r,\mu)$ with $\mu=2m/r$, the coefficient
	$B(r)$ is generally nonzero because $\delta\mu$ contains
	metric and displacement contributions.
	Thus the replacement
	$\delta p_t=(p_t'/\rho')\delta\rho$ is an additional
	assumption, not an identity.
	
	\begin{proposition}[Bulk closure nonuniqueness]
		\label{prop:bulk_nonunique}
		For anisotropic stellar matter, the equilibrium background
		profiles do not in general determine a unique interior tidal
		perturbation system. Distinct local linear closures
		$(A_1,B_1)$ and $(A_2,B_2)$ consistent with the same
		equilibrium background generally lead to different coefficients
		in the linearized bulk tidal equations and therefore need not
		produce the same interior-side logarithmic derivative
		$y_-=RH'_-/H$ at the surface.
	\end{proposition}
	
	\begin{proof}
		Equation~\eqref{eq:pertfield} contains $\delta T$ and
		$\delta\avgP$, and hence depends independently on
		$\delta p_r$, $\delta p_t$, and the metric perturbation.
		Once Eq.~\eqref{eq:closure} is imposed, the coefficients
		of the coupled first-order tidal system, or equivalently of
		its reduced second-order form, depend explicitly on the
		chosen functions $A(r)$ and $B(r)$.
		If two admissible closures differ on a set of nonzero
		measure, then the corresponding linearized bulk systems
		are generally different. The regular interior solution,
		and therefore the resulting value of $H'(R)/H(R)$,
		need not be the same.
		
		This establishes that the equilibrium background alone
		does not uniquely determine the bulk tidal closure.
		It does not assert that every distinct closure must lead
		to a numerically different $y_-$ in every special case,
		but it does show that uniqueness cannot be inferred from
		the background solution alone.
	\end{proof}
	
	\subsection{Perturbed shell}
	
	The exterior quadrupolar Love-number formula requires the
	logarithmic derivative on the Schwarzschild side of the
	stellar surface,
	\begin{equation}
		y_+
		=
		\left.\frac{rH'}{H}\right|_{R^+},
		\label{eq:y_plus_shell}
	\end{equation}
	whereas an interior tidal integration, once a bulk closure
	has been specified, yields
	\begin{equation}
		y_-
		=
		\left.\frac{rH'}{H}\right|_{R^-}.
		\label{eq:y_minus_shell}
	\end{equation}
	For a smooth perfect-fluid surface these quantities may
	coincide, apart from any known density-jump correction.
	For a thin-shell completion, however, the perturbed junction
	conditions must also involve the perturbation of the effective
	surface stress tensor introduced in Sec.~\ref{sec:shell}.
	
	The background matching fixes the equilibrium shell quantities
	$\sigma_s$ and $\mathcal P_s$, but it does not by itself determine
	their perturbations $\delta\sigma_s$ and $\delta\mathcal P_s$,
	nor the surface displacement $\xi_s$.
	Schematically, the perturbed junction system may therefore
	be expressed as
	\begin{equation}
		y_+-y_-
		=
		\mathcal J\!\left[
		C;\sigma_s,\mathcal P_s,
		\delta\sigma_s,\delta\mathcal P_s,\xi_s
		\right],
		\label{eq:jumpgeneral}
	\end{equation}
	where $\mathcal J$ denotes the relation obtained after imposing
	the perturbed geometric matching together with whatever surface
	constitutive and displacement laws are adopted.
	Equation~\eqref{eq:jumpgeneral} is schematic: it indicates
	the dependency structure of the perturbed junction problem,
	rather than providing a derived closed-form jump law for an
	arbitrary physical shell.
	
	\begin{theorem}[Tidal nonuniqueness of an unspecified thin-shell completion]
		\label{thm:tidalnonunique}
		For an equilibrium configuration that requires the effective
		surface layer of Theorem~\ref{thm:shell}, the background
		geometry and equilibrium surface stress do not by themselves
		determine a unique quadrupolar Love number.
		Even after a bulk constitutive closure has fixed $y_-$,
		the exterior-side quantity $y_+$ remains undetermined until
		the perturbed surface-response prescription is supplied.
	\end{theorem}
	
	\begin{proof}
		The Schwarzschild exterior Love number is given by the standard
		function $k_2(C,y_+)$ quoted in Appendix~\ref{app:love}.
		For a fixed compactness $C$, a unique Love number therefore
		requires a definite value of $y_+$.
		
		A specified bulk constitutive model can determine $y_-$,
		but the equilibrium Israel matching determines only the
		background shell stress. It does not supply the perturbative
		relations needed to determine $\delta\sigma_s$,
		$\delta\mathcal P_s$, or $\xi_s$, and hence it does not
		close the perturbed junction problem that links $y_-$ to $y_+$.
		
		Where $\partial k_2/\partial y_+\neq0$, different admissible
		choices of perturbed shell response may lead to different
		values of $y_+$ and hence different conditional Love numbers.
		This establishes the need for additional perturbed shell data.
		It does not claim that every conceivable shell law must yield
		a different result, only that uniqueness does not follow from
		the stated equilibrium information alone.
	\end{proof}
	
	To illustrate the sensitivity to the missing perturbed shell
	data, we introduce the conditional jump
	\begin{equation}
		\Delta y
		\equiv
		y_+-y_-.
		\label{eq:conditional_shell_jump}
	\end{equation}
	For a specified input pair $(C,y_-)$, we then evaluate
	$k_2(C,y_-+\Delta y)$ and the corresponding dimensionless
	deformability
	\begin{equation}
		\Lambda
		=
		\frac{2k_2}{3C^5}
	\end{equation}
	using the exterior formula summarized in Appendix~\ref{app:love}.
	The quantity $\Delta y$ is introduced only as an illustrative
	parameter; it is not derived here from a physical perturbed
	shell equation of state.
	
	The values of $C$ and the nominal $y_-$ used in this
	illustration are archived plotting inputs.
	In particular, Table~\ref{tab:jumps} uses
	$C=0.2451$ and $y_-=1.7242$ for PSR J0740+6620.
	The archived value $y_-=1.7242$ has not been established
	here as the output of a documented bulk tidal integration
	with a fully specified anisotropic response law.
	Accordingly, neither the zero-jump value nor the variations
	shown below should be interpreted as uniquely predicted
	stellar tidal deformabilities.
	
\begin{figure}[t]
		\centering
		\includegraphics[width=0.78\textwidth]
		{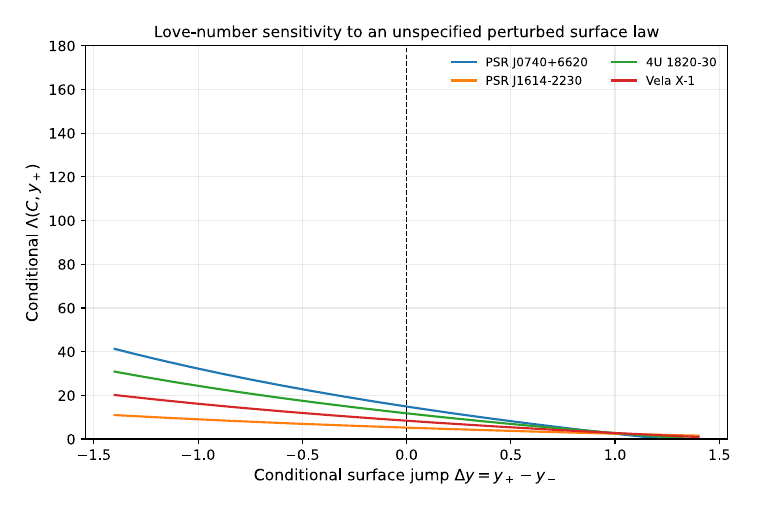}
		\caption{
			Conditional sensitivity of the dimensionless tidal
			deformability $\Lambda$ to the assumed surface jump
			$\Delta y=y_+-y_-$.
			The curves use archived compactnesses and nominal
			interior-side logarithmic-derivative inputs.
			No perturbed shell constitutive law is imposed;
			the curves therefore illustrate mathematical sensitivity
			rather than physical predictions for the named objects.
		}
		\label{fig:jump}
	\end{figure}

	Figure~\ref{fig:jump} presents the corresponding conditional
	sensitivity curves. Their purpose is to demonstrate why the
	missing perturbed shell prescription cannot be ignored even
	when the equilibrium background and the background shell
	stress are known.
	
	\begin{table}[t]
		\caption{
			Illustrative conditional tidal sensitivity for
			PSR J0740+6620, using the archived input
			$C=0.2451$ and nominal $y_-=1.7242$.
			The Love number is evaluated from
			Appendix~\ref{app:love} with
			$y_+=y_-+\Delta y$.
			The zero-jump row is a mathematical reference value,
			not a verified prediction from a specified bulk closure
			and physical perturbed shell-response model.
		}
		\label{tab:jumps}
		\centering
		\begin{tabular}{rrrr}
			\toprule
			$\Delta y$ & $y_+$ & $k_2$ & $\Lambda$ \\
			\midrule
			$-1.0$ & $0.7242$ & $0.04277022$ & $32.23545$ \\
			$-0.5$ & $1.2242$ & $0.03022138$ & $22.77753$ \\
			$ 0.0$ & $1.7242$ & $0.01974581$ & $14.88220$ \\
			$+0.5$ & $2.2242$ & $0.01086890$ & $ 8.19177$ \\
			$+1.0$ & $2.7242$ & $0.00325065$ & $ 2.44998$ \\
			\bottomrule
		\end{tabular}
	\end{table}

	\FloatBarrier
	
	\section{Verification strategy}
	\label{sec:verification}
	
	Table~\ref{tab:verification} summarizes the algebraic
	consistency checks, numerical reproductions, and
	conditional tidal-sensitivity evaluations.
	These checks apply under the adopted matter-variation
	assumption and pressure-type matter prescription.
	They establish internal consistency within these
	assumptions, but do not independently establish the
	observational provenance of the calibration inputs,
	physical admissibility, or dynamical stability.
	In particular, for positive physical surface density,
	the finite-density common-surface condition under the
	average-pressure prescription requires
	$p_t(R)=3\rho(R)/2$, violating the tangential dominant
	energy condition.
	The tidal calculations use prescribed interior-side
	inputs and assumed surface jumps; they do not constitute
	an independent integration of the bulk anisotropic
	perturbation equations or a solution of the perturbed
	shell-matching problem.
	
	\begin{table}[!htbp]
		\caption{
			Analytic consistency checks, numerical reproductions,
			and conditional sensitivity evaluations.
			The final column states the result and scope
			of each check.
		}
		\label{tab:verification}
		\centering
		\small
		\begin{tabularx}{0.96\textwidth}
			{p{0.23\textwidth}p{0.17\textwidth}X}
			\toprule
			Check & Type & Result and scope \\
			\midrule
			
			Physical--effective inverse map
			& Algebraic
			& Matrix inversion identifies the singular
			couplings $\chi=-2\pi$ and $\chi=-4\pi$
			for the adopted average-pressure prescription.
			The inverse transformation is used only
			away from these values. \\
			
			Common-surface ratio
			& Algebraic
			& Imposing $p_r=p_{r,\eff}=0$ reproduces
			the surface ratio for the general matter
			prescription, subject to the stated
			nonsingularity assumptions. \\
			
			Boundary-slope condition
			& Analytic
			& Substitution into the effective equilibrium
			equation recovers the local surface-slope
			relation and the associated necessary
			compactness inequality. \\
			
			Tangential effective pressure
			& Analytic
			& The angular Einstein equation provides
			the geometric expression for
			$p_{t,\eff}$ within the prescribed
			pseudo-spheroidal construction. \\
			
			Surface-selected coupling
			& Algebraic
			& Elimination of the material surface
			condition gives $\chi_\partial(k,s;a,p_0)$,
			its $s=1$ reduction, and the stated
			parameter specialization. \\
			
			Four-star coupling reproduction
			& Numerical
			& Substitution of archived $k$ values
			into the exact coupling relation
			reproduces the tabulated couplings
			with maximum absolute residual
			$2.55\times10^{-7}$. \\
			
			Fixed-$\chi$ compactness classes
			& \shortstack[l]{Algebraic/\\numerical}
			& The surface equation is polynomial
			in $k$ of degree at most four.
			Its roots require separate screening
			for singularities and physical
			admissibility. \\
			
			Common-surface threshold
			& Analytic
			& The positive-$p_0$ common-surface
			branch requires $C>3/7$,
			consistent with the necessary
			boundary-slope condition.
			This does not establish physical
			admissibility. \\
			
			Effective thin-shell stress
			& Analytic
			& The extrinsic-curvature jump and
			interior effective radial traction
			reproduce the same background
			surface-pressure relation.
			This effective stress does not specify
			a physical shell-response law. \\
			
			Conditional tidal sensitivity
			& Numerical
			& The exterior formula $k_2(C,y_+)$
			is evaluated using
			$y_+=y_-+\Delta y$.
			The archived $y_-$ inputs are not
			independently validated interior
			tidal solutions; the results remain
			conditional sensitivity illustrations. \\
			
			\bottomrule
		\end{tabularx}
	\end{table}
	
	\FloatBarrier

	\section{Discussion}
	\label{sec:discussion}

	\subsection{What has and has not been proved}
	
	The present analysis identifies restrictions arising
	from the combined geometric, matter, and boundary
	prescriptions rather than from the pseudo-spheroidal
	geometry itself. The metric in
	Eq.~\eqref{eq:pseudometric} is regular at the centre
	for $k>1$ and has been widely used in compact-star
	modelling
	\cite{VaidyaTikekar1982,TikekarThomas1998}.
	
	For the archived specialization, the relevant
	assumptions can be summarized as
	\begin{equation}
		\begin{aligned}
			&\text{pseudo-spheroidal }g_{rr}
			+\text{ prescribed }p_{r,\eff}\\
			&\quad+\text{ fixed }(a,p_0,s)
			+\text{ material condition }p_r(R)=0.
		\end{aligned}
		\label{eq:discussion_combined_assumptions}
	\end{equation}
	
	The physical-to-effective mapping additionally
	depends on the phenomenological matter-variation
	prescription specified in
	Sec.~\ref{sec:assumptions}.
	For $s=1$ and fixed $(a,p_0)$, the material
	boundary condition selects the coupling through
	\begin{equation}
		\chi=\chi_\partial(C).
		\label{eq:discussion_chiC}
	\end{equation}
	
	Thus, coupling--compactness locking is a consequence
	of the adopted closure rather than a general property
	of pseudo-spheroidal geometry. Changing the scale
	ratio, pressure-profile parameters, or matter
	prescription may modify the compatible configurations.
	Introducing an effective surface layer changes the
	matching conditions but does not remove the coupling
	relation imposed by $p_r(R)=0$ at fixed closure.
	
	Independently of the interior metric ansatz, a smooth
	finite-density common material/effective surface
	satisfying the assumptions of
	Theorem~\ref{thm:slope} requires, for the
	average-pressure prescription,
	\begin{equation}
		C\geq\frac{3}{7}.
	\end{equation}
	
	The same common-surface condition imposes
	\begin{equation}
		p_t(R)=\frac{3}{2}\rho(R).
	\end{equation}
	
	For $\rho(R)>0$, this contradicts the physical
	tangential dominant energy condition,
	\begin{equation}
		\rho(R)\geq |p_t(R)|.
	\end{equation}
	
	Consequently, the compactness inequality is a
	necessary local boundary-compatibility condition,
	not a sufficient criterion for physical
	admissibility. Imposing the tangential dominant
	energy condition excludes the positive-density
	common-surface configuration under the adopted
	prescription.
	
	When nonzero effective radial traction requires
	a surface layer, the resulting Israel stress
	belongs to the effective Einstein-side description
	and cannot automatically be identified with a
	physical matter-shell stress tensor. Moreover,
	the background matching does not establish
	dynamical stability or determine a unique tidal
	response without additional perturbative
	constitutive information.

	\subsection{Implications for parameter inference}
	
	A coupling value $\chi_i$ obtained for an individual
	stellar configuration by imposing the material
	boundary condition
	\begin{equation}
		p_r(R)=0
	\end{equation}
	after adopting its mass and radius $(M,R)$
	represents an inverse calibration within the
	prescribed model, rather than an independent
	observational determination of the gravitational
	coupling.
	
	The four archived coupling values are reproduced
	by the exact relation $\chi_\partial(k)$ with a
	maximum absolute residual of
	$2.55\times10^{-7}$. This agreement verifies
	the algebraic reconstruction but does not establish
	four independent measurements of a universal
	coupling. Such an interpretation would require
	a common observational likelihood, measurement
	uncertainties, and a statistical consistency
	analysis within a fixed gravitational theory
	and matter model. Since the observational
	provenance and uncertainty prescriptions of
	the adopted mass--radius inputs are not
	established, no observational confidence
	intervals are assigned to the reconstructed
	couplings.
	
	Conversely, the inability of a restricted
	metric-first closure to accommodate several
	stellar compactnesses at fixed $\chi$ does
	not constitute evidence against the underlying
	gravity theory. It identifies a limitation
	of the particular geometric and matter
	prescriptions adopted in the construction.
	
	A forward stellar model should instead specify
	the gravitational coupling, matter prescription,
	and constitutive relations before determining
	the stellar mass and radius from the equilibrium
	equations. These predictions can then be
	compared with observations
	\cite{DouchinHaensel2001,HaenselPotekhin2004,
		Read2009,LattimerPrakash2016,Yusmantoro2026}.
	Exact analytic constructions may also be
	examined within this framework when their
	remaining degrees of freedom are consistently
	specified. Varying only the dimensional scale
	at fixed dimensionless parameters produces
	a homologous constant-compactness family,
	not an independent equation-of-state sequence.
	
	\subsection{Implications for tidal calculations}
	
	The static quadrupolar Love number characterizes
	the tidal response of a stellar configuration
	and cannot, in general, be determined from its
	equilibrium metric alone. Its calculation requires
	a consistent perturbative description of the
	matter and any internal or surface interfaces
	\cite{Postnikov2010,LandryPoisson2015,
		Pani2015,BiswasBose2019,LauLeungLin2019,
		TakatsyKovacs2020,Gittins2020,
		ArangurenVera2024,Andersson2026}.
	
	For anisotropic matter, the equilibrium profiles
	$\rho(r)$, $p_r(r)$, and $p_t(r)$ do not uniquely
	specify the perturbations of the radial and
	tangential stresses. An additional constitutive
	response law is therefore required to close
	the bulk tidal perturbation equations.
	
	When an effective surface layer is present,
	Theorem~\ref{thm:tidalnonunique} establishes
	that the background Israel matching conditions
	determine the equilibrium surface stress but
	not its perturbative response. Consequently,
	the relation between the interior-side and
	exterior-side logarithmic derivatives,
	\begin{equation}
		y_-=
		\left.\frac{rH'}{H}\right|_{R^-},
		\qquad
		y_+=
		\left.\frac{rH'}{H}\right|_{R^+},
	\end{equation}
	requires appropriate perturbed junction
	conditions and a surface-response prescription.
	
	In particular, the equilibrium calculation
	does not determine a physical value of
	\begin{equation}
		\Delta y=y_+-y_-.
	\end{equation}
	
	Accordingly, the numerical results in
	Fig.~\ref{fig:jump} and
	Table~\ref{tab:jumps} represent conditional
	sensitivity calculations for assumed values
	of $\Delta y$. They demonstrate how the
	exterior Love number $k_2$ and dimensionless
	tidal deformability $\Lambda$ depend on
	the unspecified surface response.
	
	Neither the reference choice $\Delta y=0$
	nor the imposed nonzero jumps are identified
	with a verified physical matching law.
	The resulting values therefore cannot be
	interpreted as independently established
	tidal predictions for the named stellar
	configurations.
	
	A complete tidal calculation would require
	a specified bulk anisotropic constitutive law,
	regular interior perturbation equations,
	the appropriate surface classification,
	and, where a shell is present, a perturbed
	surface-stress and displacement prescription.
	Dynamical tidal-response calculations introduce
	further response information
	\cite{Hegade2024}.
	
	Thus, the principal implication of the
	present analysis is that the equilibrium
	geometry and background matching conditions
	are insufficient to determine a unique
	physical tidal response without the
	additional perturbative constitutive data.

	\subsection{Relation to stability diagnostics}
	
	The present analysis establishes equilibrium and
	boundary consistency but does not demonstrate
	radial dynamical stability. An equilibrium solution
	satisfies the corresponding field equations by
	construction; therefore, a vanishing equilibrium
	residual is not an independent stability criterion.
	
	Local diagnostics, including the sound-speed
	difference
	\begin{equation}
		v_t^2-v_r^2,
		\label{eq:discussion_sound_speed_difference}
	\end{equation}
	and the adiabatic index, may provide useful
	information about the equilibrium configuration.
	However, they cannot replace a properly formulated
	radial pulsation eigenvalue problem
	\cite{Herrera1992,Abreu2007,
		Chandrasekhar1964a,Gondek1997,
		KokkotasRuoff2001,Horvat2011,
		PretelMota2024}.
	
	For anisotropic matter, such an analysis requires
	a specified perturbative constitutive response.
	When a surface layer or internal interface is
	present, the corresponding perturbative matching
	conditions must also be imposed. Consequently,
	the background consistency checks performed
	here do not establish dynamical stability.

	\section{Conclusions}
	\label{sec:conclusion}
	
	We have investigated boundary compatibility,
	coupling--compactness locking, and tidal-response
	nonuniqueness in metric-first anisotropic stellar
	models within linear $f(R,T)$ gravity. The results
	are conditional on the adopted matter-variation
	prescription, physical-to-effective mapping,
	geometric closure, and boundary conditions.
	
	First, the material and effective-fluid surfaces
	do not generally coincide. For a smooth,
	finite-density common surface, the effective
	equilibrium equation imposes a necessary local
	compactness condition independent of the
	specific interior geometry. For the
	average-pressure matter prescription,
	this condition becomes
	\begin{equation}
		C\geq\frac{3}{7}.
		\label{eq:conclusion_compactness_bound}
	\end{equation}
	However, the same surface prescription requires
	$p_t(R)=3\rho(R)/2$, which violates the physical
	tangential dominant energy condition when
	$\rho(R)>0$. Thus, the compactness bound is
	a necessary first-order boundary-compatibility
	condition under the stated assumptions,
	not a sufficient criterion for physical
	admissibility or a universal bound on
	anisotropic compact stars.
	
	Second, when the metric is continuous across
	the stellar surface but
	\begin{equation}
		p_{r,\eff}(R)\neq0,
		\label{eq:conclusion_nonzero_preff}
	\end{equation}
	the background matching requires an effective
	Israel surface layer. For the continuous-metric
	construction considered here, the effective
	surface energy density vanishes, while the
	surface pressure satisfies
	\begin{equation}
		\mathcal{P}_s
		=
		-\frac{R\,p_{r,\eff}(R)}
		{2\sqrt{1-2C}}.
	\end{equation}
	Positive effective radial traction therefore
	produces negative surface pressure, corresponding
	to effective surface tension. The resulting
	effective layer violates the surface null,
	weak, and dominant energy conditions. These
	statements concern the Einstein-side effective
	junction stress and cannot automatically be
	transferred to a physical matter shell without
	an additional surface matter-variation
	prescription.
	
	Third, for the pseudo-spheroidal geometry
	with the prescribed effective radial-pressure
	profile, the material boundary condition
	selects the gravitational coupling through
	\begin{equation}
		\chi_\partial
		=
		\chi_\partial(k,s;a,p_0).
		\label{eq:conclusion_chi_boundary}
	\end{equation}
	For $s=1$, $a=0.8$, and $p_0=1.5$, the
	analytical relation reproduces the four
	archived coupling values with a maximum
	absolute residual of $2.55\times10^{-7}$.
	These values represent boundary-selected
	calibrations rather than independent
	observational measurements of a universal
	gravitational coupling.
	
	Conversely, at fixed nonzero $\chi$ and
	fixed dimensionless closure parameters,
	the surface equation has degree at most
	four in $k$, restricting the number of
	algebraically compatible compactness
	classes. Varying only the dimensional
	scale $L$ produces a homologous family
	with constant compactness and identical
	normalized profiles, rather than an
	independent equation-of-state-generated
	stellar sequence.
	
	Finally, the equilibrium geometry and
	background surface stress do not uniquely
	determine the quadrupolar tidal Love number.
	A complete physical calculation requires
	a specified bulk anisotropic perturbation
	response and, when a surface layer is
	present, appropriate perturbed junction
	conditions and surface constitutive relations.
	The numerical Love-number and tidal-
	deformability values reported here are
	conditional sensitivity calculations based
	on archived interior-side inputs and
	assumed surface jumps. They therefore do
	not constitute independently established
	physical tidal predictions.
	
	Overall, the analysis demonstrates that
	boundary conditions and constitutive
	assumptions play a central role in the
	interpretation of metric-first stellar
	constructions. The identified restrictions
	arise from the combined geometric, matter,
	and boundary prescriptions rather than
	from an intrinsic inconsistency of
	pseudo-spheroidal geometry or linear
	$f(R,T)$ gravity. The resulting analytical
	relations provide consistency tests for
	distinguishing model-dependent boundary
	calibrations from genuine gravitational
	theory parameters and for identifying
	the additional information required for
	physical tidal and stability analyses.
	
	\appendix
	
	\section{Full polynomial in the surface-selected coupling}
	\label{app:poly}
	
	The exact surface-selected coupling obtained from the
	material boundary condition is
	\begin{equation}
		\chi_\partial(k,s;a,p_0)
		=
		\frac{
			48\pi p_0(1+s^2)^2
			(1-as^2)(1+ks^2)
		}{
			\calD(k,s;a,p_0)
		}.
	\end{equation}
	
	The denominator is the dimensionless polynomial
	\begin{align}
		\calD={}&
		a^2p_0^2s^{10}
		+2a^2p_0^2s^8
		+a^2p_0^2s^6
		-2ak^2p_0s^8
		-2ak^2p_0s^6
		\nonumber\\
		&+20akp_0s^8
		+38akp_0s^6
		+18akp_0s^4
		-2ap_0^2s^8
		-4ap_0^2s^6
		-2ap_0^2s^4
		\nonumber\\
		&+14ap_0s^6
		+28ap_0s^4
		+14ap_0s^2
		+k^4s^6
		-8k^3s^6
		-2k^3s^4
		\nonumber\\
		&+2k^2p_0s^6
		+2k^2p_0s^4
		+7k^2s^6
		-26k^2s^4
		-21k^2s^2
		\nonumber\\
		&-24kp_0s^6
		-46kp_0s^4
		-22kp_0s^2
		+28ks^4
		-18k
		\nonumber\\
		&+p_0^2s^6
		+2p_0^2s^4
		+p_0^2s^2
		-18p_0s^4
		-36p_0s^2
		-18p_0
		+21s^2
		+18.
		\label{eq:Dfull}
	\end{align}
	
	The polynomial is of degree four in the spheroidicity
	parameter $k$ for $s>0$.
	It is constructed from the prescribed pseudo-spheroidal
	metric, the effective radial-pressure profile, and the
	angular Einstein equation used to determine the
	effective tangential pressure.
	
	As an algebraic consistency check, setting $s=1$ gives
	\begin{align}
		\calD(k,1;a,p_0)
		={}&
		k^4-10k^3
		+\left[4p_0(1-a)-40\right]k^2
		\nonumber\\
		&+\left[p_0(76a-92)+10\right]k
		\nonumber\\
		&+4p_0^2(a-1)^2
		+p_0(56a-72)+39.
	\end{align}
	
	For the parameter specialization
	\begin{equation}
		a=\frac45,
		\qquad
		p_0=\frac32,
		\qquad
		s=1,
	\end{equation}
	the denominator reduces to
	\begin{equation}
		\calD
		=
		\frac{1}{25}
		\left(
		25k^4-250k^3-970k^2
		-920k-36
		\right).
	\end{equation}
	
	Consequently, the specialized surface-selected
	relation is
	\begin{equation}
		\chi_\partial(k)
		=
		\frac{
			1440\pi(k+1)
		}{
			25k^4-250k^3-970k^2
			-920k-36
		}.
	\end{equation}
	
	These reductions agree algebraically with the
	surface-selected coupling used in the main text.
	
	The zeros of $\calD$ correspond to poles of the
	explicitly solved expression for $\chi_\partial$
	when its numerator remains nonzero.
	At such points, the original material surface equation
	admits no finite coupling through the displayed
	solved branch.
	
	If both numerator and denominator vanish simultaneously,
	the explicit ratio becomes indeterminate.
	Such points must be examined using the original
	unsolved surface equation rather than by direct
	substitution into the ratio for $\chi_\partial$.
	
	A real algebraic root does not by itself establish
	a physically acceptable stellar configuration.
	The restrictions on $k$ and $s$, regularity,
	the physical-to-effective inverse map, and the
	selected boundary class must be applied separately.

	\section{Love-number expression and conditional jump map}
	\label{app:love}
	
	For a Schwarzschild vacuum exterior, we denote by
	\begin{equation}
		y_+
		=
		\left.
		\frac{rH'(r)}{H(r)}
		\right|_{R^+}
	\end{equation}
	the exterior-side logarithmic derivative of the static
	quadrupolar metric perturbation, where $H(R^+)\neq0$.
	
	For a stellar compactness
	\begin{equation}
		C=\frac{M}{R},
		\qquad
		0<C<\frac12,
	\end{equation}
	the standard quadrupolar exterior Love-number expression is
	\begin{equation}
		k_2
		=
		\frac{8C^5}{5}(1-2C)^2
		\frac{
			2+2C(y_+-1)-y_+
		}{
			\mathcal Z(C,y_+)
		},
		\label{eq:love_k2}
	\end{equation}
	where
	\begin{align}
		\mathcal Z(C,y_+)
		={}&
		2C\left[
		6-3y_+
		+3C(5y_+-8)
		\right]
		\nonumber\\
		&+
		4C^3
		\left[
		13-11y_+
		+C(3y_+-2)
		+2C^2(1+y_+)
		\right]
		\nonumber\\
		&+
		3(1-2C)^2
		\left[
		2-y_+
		+2C(y_+-1)
		\right]
		\ln(1-2C).
		\label{eq:love_Z}
	\end{align}
	
	The expression is evaluated where
	$\mathcal Z(C,y_+)\neq0$.
	The corresponding dimensionless quadrupolar tidal
	deformability is
	\begin{equation}
		\Love
		=
		\frac{2k_2}{3C^5}.
		\label{eq:tidal_deformability}
	\end{equation}
	
	The exterior Love-number expression requires the
	exterior-side value $y_+$.
	For a smooth fluid boundary, its relation to the
	interior-side value $y_-$ follows from the applicable
	perturbative matching conditions.
	When an effective thin shell is present, the background
	junction conditions alone do not determine this relation.
	The perturbed surface stress and displacement must also
	be specified before the exterior value can be calculated
	from a physical interior perturbation solution.
	
	To examine this dependence without assuming a particular
	shell constitutive law, the conditional sensitivity
	analysis uses the substitution
	\begin{equation}
		y_+\longrightarrow y_-+\Delta y,
		\label{eq:conditional_jump_map}
	\end{equation}
	where
	\begin{equation}
		\Delta y=y_+-y_-.
	\end{equation}
	For each adopted pair $(C,y_-)$, the quantities
	\begin{equation}
		k_2(C,y_-+\Delta y),
		\qquad
		\Love(C,y_-+\Delta y)
	\end{equation}
	are evaluated while varying $\Delta y$ as an
	illustrative parameter.
	
	The adopted input pairs are
	\begin{align}
		(C,y_-)_{\mathrm{J0740}}
		&=(0.2451,1.7242),
		\nonumber\\
		(C,y_-)_{\mathrm{J1614}}
		&=(0.2999,1.3008),
		\nonumber\\
		(C,y_-)_{\mathrm{4U1820}}
		&=(0.2561,1.6801),
		\nonumber\\
		(C,y_-)_{\mathrm{Vela}}
		&=(0.2731,1.5817).
	\end{align}
	The nominal interior-side values $y_-$ are prescribed
	illustrative inputs rather than results of a documented
	bulk anisotropic tidal perturbation integration.
	Their interpretation and the limitations of the
	calibration inputs are discussed in
	Section~\ref{sec:fourstars}.
	
	The conditional sensitivity results are calculated
	using Eqs.~\eqref{eq:love_k2}--\eqref{eq:tidal_deformability}
	with the substitution in
	Eq.~\eqref{eq:conditional_jump_map}.
	For example, using the adopted PSR J0740+6620 input
	and $\Delta y=0$ gives
	\begin{equation}
		y_+=1.7242,
		\qquad
		k_2\simeq0.01974581,
	\end{equation}
	and
	\begin{equation}
		\Love\simeq14.88220.
	\end{equation}
	
	These values reproduce the zero-jump row of
	Table~\ref{tab:jumps}.
	The remaining entries are calculated using the same
	exterior Love-number expression for the prescribed
	nonzero values of $\Delta y$.
	
	Accordingly, Fig.~\ref{fig:jump} and
	Table~\ref{tab:jumps} illustrate the mathematical
	sensitivity of $k_2$ and $\Love$ to the assumed
	surface jump.
	They do not represent independently established
	physical tidal predictions.
	
	\section{Boundary classes and required data}
	\label{app:classes}
	
	The classification below summarizes
	the background matching conditions
	and additional constitutive information
	required for the different stellar
	boundary constructions considered
	in this work.
	
	A dilute smooth surface is
	characterized here by the vanishing
	of the relevant physical matter
	variables at the stellar boundary.
	
	A smooth finite-density common
	material/effective surface must
	satisfy the algebraic common-surface
	relation and the appropriate
	boundary-slope condition, in
	addition to the geometric matching
	requirements.
	
	For the average-pressure prescription,
	a positive-density common surface
	requires $p_t(R)=3\rho(R)/2$ and
	therefore violates the physical
	tangential dominant energy condition.
	
	A continuous-metric thin-shell
	completion is described by an
	effective surface stress determined
	from the extrinsic-curvature jump.
	
	Its perturbative response requires
	additional constitutive information
	that is not provided by the
	background geometry.
	
	The effective surface layer must
	also be distinguished from a
	physical distributional matter
	shell in the underlying $f(R,T)$
	theory.
	
	Sharp internal interfaces require
	appropriate continuity of the
	induced geometry and normal
	traction, together with
	thermodynamic and perturbative
	interface prescriptions.
	
	Finally, a homologous scale family
	is not a separate boundary class.
	It is included in the following
	table only to distinguish scaling
	freedom from the choice of
	stellar surface conditions.
	
	{\small
		\sloppy
		\setlength{\tabcolsep}{3pt}
		\begin{longtable}{
				>{\raggedright\arraybackslash}p{0.15\textwidth}
				>{\raggedright\arraybackslash}p{0.20\textwidth}
				>{\raggedright\arraybackslash}p{0.22\textwidth}
				>{\raggedright\arraybackslash}p{0.25\textwidth}
			}
			
			\caption{
				Logical classification of stellar
				boundary prescriptions and the
				additional information required
				for background and tidal calculations.
				The homologous scale family is
				included for comparison and is
				not itself a boundary class.
			}
			\label{tab:classes}\\
			
			\toprule
			Boundary class
			& Background conditions
			& Additional checks
			& Tidal data required \\
			\midrule
			\endfirsthead
			
			\caption[]{Boundary classes and required data
				(continued).}\\
			
			\toprule
			Boundary class
			& Background conditions
			& Additional checks
			& Tidal data required \\
			\midrule
			\endhead
			
			\midrule
			\multicolumn{4}{r}{
				\small Continued on next page
			}\\
			\endfoot
			
			\bottomrule
			\endlastfoot
			
			Dilute smooth surface
			&
			$p_r=p_{r,\eff}=\rho=p_t=0$
			at the boundary in the
			dilute case considered here
			&
			Central regularity, effective
			equilibrium, monotonicity,
			and smooth geometric matching
			&
			Bulk constitutive response,
			regular tidal integration,
			and smooth-surface matching
			\\
			
			Finite-density smooth surface
			&
			$p_r=p_{r,\eff}=0$,
			$\rho>0$, with continuous
			extrinsic curvature
			&
			Common-surface ratio,
			boundary-slope condition,
			physical and effective
			energy conditions, and
			junction consistency
			&
			Bulk perturbation closure
			and the appropriate
			finite-density surface
			matching or jump relation
			\\
			
			Thin-shell surface
			&
			$p_r=0$,
			$p_{r,\eff}\neq0$,
			with a continuous
			induced metric
			&
			Effective Israel stress,
			surface energy conditions
			if imposed, and distinction
			between effective and
			physical shell matter
			&
			Bulk closure,
			perturbed surface stress
			law, displacement
			prescription, and
			perturbed junction
			conditions
			\\
			
			Sharp internal interface
			&
			Continuity of the appropriate
			normal traction, with a
			possible density or
			constitutive discontinuity
			&
			Interface thermodynamics,
			geometric matching,
			and traction balance
			&
			Interface displacement
			or conversion law and
			the corresponding
			metric-perturbation
			matching conditions
			\\
			
			Homologous scale family
			(comparison)
			&
			Fixed dimensionless
			parameters with variable
			dimensional scale $L$
			&
			Constant compactness,
			invariant normalized
			profiles, and consistent
			dimensional scaling
			&
			Not a common-EOS stellar
			sequence without an
			independently specified
			dimensional constitutive law
			\\
			
		\end{longtable}
	}
	
	A homologous scale family changes
	the dimensional mass and radius
	while keeping the compactness
	and normalized profiles fixed.
	
	This scaling behaviour does not
	replace the need for a physically
	consistent boundary prescription,
	nor does it define a unique
	tidal response.
	
	The bulk and surface perturbation
	data must therefore be specified
	independently of the dimensional
	scale transformation.

\section*{Data availability}

The data and computational materials supporting the findings of this study are available from the corresponding authors upon reasonable request.

\section*{Statements and Declarations}

\subsection*{Funding}

The authors declare that no specific funding was received for this work.

\subsection*{Competing interests}

The authors declare no competing interests.

\subsection*{Author contributions}

H.R.A.: Investigation, symbolic and numerical implementation, visualisation, and original manuscript drafting.

D.M.P.: Conceptualisation, formal analysis, methodology, verification, supervision, and writing--review and editing.

B.B.P.: Boundary and junction analysis, validation, and writing--review and editing.

V.O.T.: Pseudo-spheroidal geometry, analytic-limit review, supervision, and writing--review and editing.

All authors reviewed and approved the final version of the manuscript.

\section*{Use of AI-assisted tools}
ChatGPT was used as an assistive tool during the preparation of this work,
primarily for code development, language revision, manuscript organisation, and
preliminary consistency checks between analytical expressions, numerical
calculations, figures, and manuscript descriptions. The authors reviewed the calculations, equations, figures, references, interpretations, and conclusions and retain full responsibility for the content of the manuscript.
	
	\printbibliography[title={References}]

@article{Tolman1939,
  author={Tolman, R. C.}, title={Static Solutions of Einstein's Field Equations for Spheres of Fluid},
  journal={Physical Review}, volume={55}, pages={364--373}, year={1939}, doi={10.1103/PhysRev.55.364}}

@article{OppenheimerVolkoff1939,
  author={Oppenheimer, J. R. and Volkoff, G. M.}, title={On Massive Neutron Cores},
  journal={Physical Review}, volume={55}, pages={374--381}, year={1939}, doi={10.1103/PhysRev.55.374}}

@article{Buchdahl1959,
  author={Buchdahl, H. A.}, title={General Relativistic Fluid Spheres},
  journal={Physical Review}, volume={116}, pages={1027--1034}, year={1959}, doi={10.1103/PhysRev.116.1027}}

@article{Chandrasekhar1964a,
  author={Chandrasekhar, S.}, title={Dynamical Instability of Gaseous Masses Approaching the Schwarzschild Limit in General Relativity},
  journal={Physical Review Letters}, volume={12}, pages={114--116}, year={1964}, doi={10.1103/PhysRevLett.12.114}}

@article{Chandrasekhar1964b,
  author={Chandrasekhar, S.}, title={The Dynamical Instability of Gaseous Masses Approaching the Schwarzschild Limit in General Relativity},
  journal={Astrophysical Journal}, volume={140}, pages={417--433}, year={1964}, doi={10.1086/147938}}

@article{BowersLiang1974,
  author={Bowers, R. L. and Liang, E. P. T.}, title={Anisotropic Spheres in General Relativity},
  journal={Astrophysical Journal}, volume={188}, pages={657--665}, year={1974}, doi={10.1086/152760}}

@article{Ruderman1972,
  author={Ruderman, M.}, title={Pulsars: Structure and Dynamics},
  journal={Annual Review of Astronomy and Astrophysics}, volume={10}, pages={427--476}, year={1972}, doi={10.1146/annurev.aa.10.090172.002235}}

@article{Herrera1992,
  author={Herrera, L.}, title={Cracking of Self-Gravitating Compact Objects},
  journal={Physics Letters A}, volume={165}, pages={206--210}, year={1992}, doi={10.1016/0375-9601(92)90036-L}}

@article{HerreraSantos1997,
  author={Herrera, L. and Santos, N. O.}, title={Local Anisotropy in Self-Gravitating Systems},
  journal={Physics Reports}, volume={286}, pages={53--130}, year={1997}, doi={10.1016/S0370-1573(96)00042-7}}

@article{Abreu2007,
  author={Abreu, H. and Hernandez, H. and Nunez, L. A.}, title={Sound Speeds, Cracking and the Stability of Self-Gravitating Anisotropic Compact Objects},
  journal={Classical and Quantum Gravity}, volume={24}, pages={4631--4646}, year={2007}, doi={10.1088/0264-9381/24/18/005}}

@article{DevGleiser2002,
  author={Dev, K. and Gleiser, M.}, title={Anisotropic Stars: Exact Solutions},
  journal={General Relativity and Gravitation}, volume={34}, pages={1793--1818}, year={2002}, doi={10.1023/A:1020707906543}}

@article{DevGleiser2003,
  author={Dev, K. and Gleiser, M.}, title={Anisotropic Stars II: Stability},
  journal={General Relativity and Gravitation}, volume={35}, pages={1435--1457}, year={2003}, doi={10.1023/A:1024534702166}}

@article{Ivanov2002,
  author={Ivanov, B. V.}, title={Maximum Bounds on the Surface Redshift of Anisotropic Stars},
  journal={Physical Review D}, volume={65}, pages={104011}, year={2002}, doi={10.1103/PhysRevD.65.104011}}

@article{DelgatyLake1998,
  author={Delgaty, M. S. R. and Lake, K.}, title={Physical Acceptability of Isolated, Static, Spherically Symmetric, Perfect Fluid Solutions of Einstein's Equations},
  journal={Computer Physics Communications}, volume={115}, pages={395--415}, year={1998}, doi={10.1016/S0010-4655(98)00130-1}}

@article{Lake2003,
  author={Lake, K.}, title={All Static Spherically Symmetric Perfect-Fluid Solutions of Einstein's Equations},
  journal={Physical Review D}, volume={67}, pages={104015}, year={2003}, doi={10.1103/PhysRevD.67.104015}}

@article{VaidyaTikekar1982,
  author={Vaidya, P. C. and Tikekar, R.}, title={Exact Relativistic Model for a Superdense Star},
  journal={Journal of Astrophysics and Astronomy}, volume={3}, pages={325--334}, year={1982}, doi={10.1007/BF02714870}}

@article{TikekarThomas1998,
  author  = {R. Tikekar and V. O. Thomas},
  title   = {Relativistic fluid sphere on pseudo-spheroidal space-time},
  journal = {Pramana},
  volume  = {50},
  pages   = {95--103},
  year    = {1998},
  doi     = {10.1007/BF02847521}
}

@article{FinchSkea1989,
  author={Finch, M. R. and Skea, J. E. F.}, title={A Realistic Stellar Model Based on an Ansatz of Duorah and Ray},
  journal={Classical and Quantum Gravity}, volume={6}, pages={467--476}, year={1989}, doi={10.1088/0264-9381/6/4/007}}

@article{KroriBarua1975,
  author  = {Krori, K. D. and Barua, J.},
  title   = {A singularity-free solution for a charged fluid
             sphere in general relativity},
  journal = {Journal of Physics A: Mathematical and General},
  volume  = {8},
  pages   = {508--511},
  year    = {1975},
  doi     = {10.1088/0305-4470/8/4/012}
}

@article{Karmarkar1948,
  author={Karmarkar, K. R.}, title={Gravitational Metrics of Spherical Symmetry and Class One},
  journal={Proceedings of the Indian Academy of Sciences A}, volume={27}, pages={56--60}, year={1948}, doi={10.1007/BF03173443}}

@article{Darmois1927,
  author={Darmois, G.}, title={Les Equations de la Gravitation Einsteinienne},
  journal={Memorial des Sciences Mathematiques}, volume={25}, pages={1--48}, year={1927}}

@article{Israel1966,
  author={Israel, W.}, title={Singular Hypersurfaces and Thin Shells in General Relativity},
  journal={Il Nuovo Cimento B}, volume={44}, pages={1--14}, year={1966}, doi={10.1007/BF02710419}}

@article{MarsSenovilla1993,
  author={Mars, M. and Senovilla, J. M. M.}, title={Geometry of General Hypersurfaces in Spacetime: Junction Conditions},
  journal={Classical and Quantum Gravity}, volume={10}, pages={1865--1897}, year={1993}, doi={10.1088/0264-9381/10/9/026}}

@article{Senovilla2013,
  author={Senovilla, J. M. M.}, title={Junction Conditions for F(R)-Gravity and Their Consequences},
  journal={Physical Review D}, volume={88}, pages={064015}, year={2013}, doi={10.1103/PhysRevD.88.064015}}

@book{Poisson2004,
  author={Poisson, E.}, title={A Relativist's Toolkit: The Mathematics of Black-Hole Mechanics},
  publisher={Cambridge University Press}, address={Cambridge}, year={2004}, doi={10.1017/CBO9780511606601}}

@article{Harko2011,
  author={Harko, T. and Lobo, F. S. N. and Nojiri, S. and Odintsov, S. D.}, title={$f(R,T)$ Gravity},
  journal={Physical Review D}, volume={84}, pages={024020}, year={2011}, doi={10.1103/PhysRevD.84.024020}}

@article{BarrientosRubilar2014,
  author={Barrientos O., J. and Rubilar, G. F.}, title={Comment on ``$f(R,T)$ Gravity''},
  journal={Physical Review D}, volume={90}, pages={028501}, year={2014}, doi={10.1103/PhysRevD.90.028501}}

@article{HarkoLobo2010,
  author={Harko, T. and Lobo, F. S. N.}, title={$f(R,L_m)$ Gravity},
  journal={European Physical Journal C}, volume={70}, pages={373--379}, year={2010}, doi={10.1140/epjc/s10052-010-1467-3}}

@article{Brown1993,
  author={Brown, J. D.}, title={Action Functionals for Relativistic Perfect Fluids},
  journal={Classical and Quantum Gravity}, volume={10}, pages={1579--1606}, year={1993}, doi={10.1088/0264-9381/10/8/017}}

@article{Faraoni2009,
  author={Faraoni, V.}, title={Lagrangian Description of Perfect Fluids and Modified Gravity with an Extra Force},
  journal={Physical Review D}, volume={80}, pages={124040}, year={2009}, doi={10.1103/PhysRevD.80.124040}}

@article{MinazzoliHarko2012,
  author={Minazzoli, O. and Harko, T.}, title={New Derivation of the Lagrangian of a Perfect Fluid with a Barotropic Equation of State},
  journal={Physical Review D}, volume={86}, pages={087502}, year={2012}, doi={10.1103/PhysRevD.86.087502}}

@article{AvelinoAzevedo2018,
  author={Avelino, P. P. and Azevedo, R. P. L.}, title={Perfect Fluid Lagrangian and Its Cosmological Implications in Theories of Gravity with Nonminimally Coupled Matter Fields},
  journal={Physical Review D}, volume={97}, pages={064018}, year={2018}, doi={10.1103/PhysRevD.97.064018}}

@article{FisherCarlson2019,
  author={Fisher, S. B. and Carlson, E. D.}, title={Reexamining $f(R,T)$ Gravity},
  journal={Physical Review D}, volume={100}, pages={064059}, year={2019}, doi={10.1103/PhysRevD.100.064059}}

@article{Moraes2016,
  author={Moraes, P. H. R. S. and Arbani, J. D. V. and Malheiro, M.}, title={Stellar Equilibrium Configurations of Compact Stars in $f(R,T)$ Gravity},
  journal={Journal of Cosmology and Astroparticle Physics}, volume={2016}, number={06}, pages={005}, year={2016}, doi={10.1088/1475-7516/2016/06/005}}

@article{Das2016,
  author={Das, A. and Rahaman, F. and Guha, B. K. and Ray, S.}, title={Compact Stars in $f(R,T)$ Gravity},
  journal={European Physical Journal C}, volume={76}, pages={654}, year={2016}, doi={10.1140/epjc/s10052-016-4503-0}}

@article{Deb2018,
  author={Deb, D. and Guha, B. K. and Rahaman, F. and Ray, S.}, title={Anisotropic Strange Stars under Simplest Minimal Matter-Geometry Coupling in the $f(R,T)$ Gravity},
  journal={Physical Review D}, volume={97}, pages={084026}, year={2018}, doi={10.1103/PhysRevD.97.084026}}

@article{Biswas2019AnnPhys,
  author={Biswas, S. and Ghosh, S. and Ray, S. and Guha, B. K. and Rahaman, F.}, title={Strange Stars in Krori-Barua Spacetime under $f(R,T)$ Gravity},
  journal={Annals of Physics}, volume={401}, pages={1--20}, year={2019}, doi={10.1016/j.aop.2018.12.018}}

@article{Dey2021,
  author={Dey, S. and Chanda, A. and Paul, B. C.}, title={Compact Objects in $f(R,T)$ Gravity with Finch-Skea Geometry},
  journal={European Physical Journal Plus}, volume={136}, pages={228}, year={2021}, doi={10.1140/epjp/s13360-021-01173-w}}

@article{RejBhar2021,
  author={Rej, P. and Bhar, P.}, title={Charged Strange Star in $f(R,T)$ Gravity with Linear Equation of State},
  journal={Astrophysics and Space Science}, volume={366}, pages={35}, year={2021}, eprint={2105.12572}, archivePrefix={arXiv}, primaryClass={gr-qc}}

@article{RejBharGovender2021,
  author={Rej, P. and Bhar, P. and Govender, M.}, title={Charged Compact Star in $f(R,T)$ Gravity in Tolman-Kuchowicz Spacetime},
  journal={European Physical Journal C}, volume={81}, pages={316}, year={2021}, eprint={2105.15130}, archivePrefix={arXiv}, primaryClass={gr-qc}}

@article{Pretel2021,
  author={Pretel, J. M. Z. and Joras, S. E. and Reis, R. R. R. and Arbanil, J. D. V.}, title={Radial Oscillations and Stability of Compact Stars in $f(R,T)=R+2\beta T$ Gravity},
  journal={Journal of Cosmology and Astroparticle Physics}, volume={2021}, number={04}, pages={064}, year={2021}, doi={10.1088/1475-7516/2021/04/064}}

@article{Rosa2021,
  author={Rosa, J. L.}, title={Junction Conditions and Thin Shells in Perfect-Fluid $f(R,T)$ Gravity},
  journal={Physical Review D}, volume={103}, pages={104069}, year={2021}, doi={10.1103/PhysRevD.103.104069}}

@article{Carvalho2021,
  author={Carvalho, G. A. and Lobato, R. V. and Deb, D. and Moraes, P. H. R. S. and Arbanil, J. D. V. and Malheiro, M.}, title={Stellar Equilibrium Configurations of White Dwarfs in $f(R,T)$ Gravity},
  journal={European Physical Journal C}, volume={81}, pages={134}, year={2021}, doi={10.1140/epjc/s10052-021-08920-4}}

@article{Chowdhury2025,
  author={Chowdhury, A. R. and Das, S. and Rahaman, F.}, title={A Study of Compact Stellar Objects in $f(R,T)$ Theory of Gravity},
  journal={Universe}, volume={11}, pages={409}, year={2025}, doi={10.3390/universe11120409}}

@article{Nazar2026,
  author={Nazar, H. and Majeed, A. and Abbas, G. and Alrashood, J. S. and Abdelfattah, W.}, title={Geometric Effects of Linear Matter-Curvature Coupling on Relativistic Stellar Structures},
  journal={International Journal of Modern Physics A}, volume={41}, pages={2650127}, year={2026}, doi={10.1142/S0217751X26501277}}

@article{Yusmantoro2026,
  author={Yusmantoro, Y. and Zen, F. P. and Pattersons, M. L.}, title={Mass-Radius Relation, Moment of Inertia and Tidal Love Numbers of Anisotropic Neutron Stars in $f(R,T)$ Gravity},
  journal={International Journal of Modern Physics D}, volume={35}, number={03}, pages={2550100}, year={2026}, doi={10.1142/S0218271825501007}}

@article{Hinderer2008,
  author={Hinderer, T.}, title={Tidal Love Numbers of Neutron Stars},
  journal={Astrophysical Journal}, volume={677}, pages={1216--1220}, year={2008}, doi={10.1086/533487}, note={Erratum: Astrophys. J. 697, 964 (2009)}}

@article{FlanaganHinderer2008,
  author={Flanagan, E. E. and Hinderer, T.}, title={Constraining Neutron-Star Tidal Love Numbers with Gravitational-Wave Detectors},
  journal={Physical Review D}, volume={77}, pages={021502}, year={2008}, doi={10.1103/PhysRevD.77.021502}}

@article{BinningtonPoisson2009,
  author={Binnington, T. and Poisson, E.}, title={Relativistic Theory of Tidal Love Numbers},
  journal={Physical Review D}, volume={80}, pages={084018}, year={2009}, doi={10.1103/PhysRevD.80.084018}}

@article{DamourNagar2009,
  author={Damour, T. and Nagar, A.}, title={Relativistic Tidal Properties of Neutron Stars},
  journal={Physical Review D}, volume={80}, pages={084035}, year={2009}, doi={10.1103/PhysRevD.80.084035}}

@article{Postnikov2010,
  author={Postnikov, S. and Prakash, M. and Lattimer, J. M.}, title={Tidal Love Numbers of Neutron and Self-Bound Quark Stars},
  journal={Physical Review D}, volume={82}, pages={024016}, year={2010}, doi={10.1103/PhysRevD.82.024016}}

@article{Hinderer2010,
  author={Hinderer, T. and Lackey, B. D. and Lang, R. N. and Read, J. S.}, title={Tidal Deformability of Neutron Stars with Realistic Equations of State and Their Gravitational Wave Signatures in Binary Inspiral},
  journal={Physical Review D}, volume={81}, pages={123016}, year={2010}, doi={10.1103/PhysRevD.81.123016}}

@article{DamourNagarVillain2012,
  author={Damour, T. and Nagar, A. and Villain, L.}, title={Measurability of the Tidal Polarizability of Neutron Stars in Late-Inspiral Gravitational-Wave Signals},
  journal={Physical Review D}, volume={85}, pages={123007}, year={2012}, doi={10.1103/PhysRevD.85.123007}}

@article{LandryPoisson2015,
  author={Landry, P. and Poisson, E.}, title={Tidal Deformation of a Slowly Rotating Material Body: External Metric},
  journal={Physical Review D}, volume={91}, pages={104018}, year={2015}, doi={10.1103/PhysRevD.91.104018}}

@article{Pani2015,
  author={Pani, P. and Gualtieri, L. and Ferrari, V.}, title={Tidal Love Numbers of a Slowly Spinning Neutron Star},
  journal={Physical Review D}, volume={92}, pages={024010}, year={2015}, doi={10.1103/PhysRevD.92.024010}}

@article{BiswasBose2019,
  author  = {Biswas, B. and Bose, S.},
  title   = {Tidal Deformability of an Anisotropic Compact Star:
             Implications of {GW170817}},
  journal = {Physical Review D},
  volume  = {99},
  pages   = {104002},
  year    = {2019},
  doi     = {10.1103/PhysRevD.99.104002},
  note    = {Erratum: Physical Review D 111, 029903 (2025),
             doi: 10.1103/PhysRevD.111.029903}
}

@article{LauLeungLin2019,
  author={Lau, S. Y. and Leung, P. T. and Lin, L. M.}, title={Tidal Deformations of Compact Stars with Crystalline Quark Matter},
  journal={Physical Review D}, volume={99}, pages={023018}, year={2019}, doi={10.1103/PhysRevD.99.023018}}

@article{TakatsyKovacs2020,
  author={Takatsy, J. and Kovacs, P.}, title={Comment on ``Tidal Love Numbers of Neutron and Self-Bound Quark Stars''},
  journal={Physical Review D}, volume={102}, pages={028501}, year={2020}, doi={10.1103/PhysRevD.102.028501}}

@article{Gittins2020,
  author={Gittins, F. and Andersson, N. and Pereira, J. P.}, title={Tidal Deformations of Neutron Stars with Elastic Crusts},
  journal={Physical Review D}, volume={101}, pages={103025}, year={2020}, doi={10.1103/PhysRevD.101.103025}}

@article{ArangurenVera2024,
  author={Aranguren, E. and Vera, R.}, title={Review on the Matching Conditions for the Tidal Problem: Towards the Application to More General Contexts},
  journal={General Relativity and Gravitation}, volume={56}, pages={50}, year={2024}, doi={10.1007/s10714-024-03237-5}}

@article{Lau2024,
  author={Lau, S. Y. and Ajith, S. and Guedes, V. and Yagi, K.}, title={Nonradial Instabilities in Anisotropic Neutron Stars},
  journal={Physical Review D}, volume={110}, pages={083020}, year={2024}, doi={10.1103/PhysRevD.110.083020}}

@article{Guedes2026,
  author={Guedes, V. and Ajith, S. and Lau, S. Y. and Yagi, K.}, title={Gravitational-Wave Constraints on Neutron-Star Pressure Anisotropy via Universal Relations},
  journal={Physical Review D}, volume={113}, pages={043025}, year={2026}, doi={10.1103/s9zp-jfnh}}

@article{Andersson2026,
  author={Andersson, N. and Counsell, A. R. and Gittins, F. and Ghosh, S.}, title={Tidal Response of a Relativistic Star},
  journal={Physical Review D}, volume={113}, pages={064051}, year={2026}, doi={10.1103/x4yr-x1c4}}

@article{Horvat2011,
  author={Horvat, D. and Ilijic, S. and Marunovic, A.}, title={Radial Pulsations and Stability of Anisotropic Stars with a Quasi-Local Equation of State},
  journal={Classical and Quantum Gravity}, volume={28}, pages={025009}, year={2011}, doi={10.1088/0264-9381/28/2/025009}}

@article{Gondek1997,
  author={Gondek, D. and Haensel, P. and Zdunik, J. L.}, title={Radial Pulsations and Stability of Protoneutron Stars},
  journal={Astronomy and Astrophysics}, volume={325}, pages={217--227}, year={1997}}

@article{KokkotasRuoff2001,
  author={Kokkotas, K. D. and Ruoff, J.}, title={Radial Oscillations of Relativistic Stars},
  journal={Astronomy and Astrophysics}, volume={366}, pages={565--572}, year={2001}, doi={10.1051/0004-6361:20000216}}

@article{PretelMota2024,
  author={Pretel, J. M. Z. and Mota, C. E.}, title={Compact Stars in Rastall Gravity: Hydrostatic Equilibrium and Radial Pulsations},
  journal={General Relativity and Gravitation}, volume={56}, pages={43}, year={2024}, doi={10.1007/s10714-024-03225-9}}

@article{Hegade2024,
    author = {Hegade K. R., Abhishek and
              Ripley, Justin L. and
              Yunes, Nicol{\'a}s},
    title = {Dynamical tidal response of
             nonrotating relativistic stars},
    journaltitle = {Physical Review D},
    volume = {109},
    number = {10},
    eid = {104064},
    year = {2024},
    doi = {10.1103/PhysRevD.109.104064}
}

@article{DouchinHaensel2001,
  author={Douchin, F. and Haensel, P.}, title={A Unified Equation of State of Dense Matter and Neutron Star Structure},
  journal={Astronomy and Astrophysics}, volume={380}, pages={151--167}, year={2001}, doi={10.1051/0004-6361:20011402}}

@article{HaenselPotekhin2004,
  author={Haensel, P. and Potekhin, A. Y.}, title={Analytical Representations of Unified Equations of State of Neutron-Star Matter},
  journal={Astronomy and Astrophysics}, volume={428}, pages={191--197}, year={2004}, doi={10.1051/0004-6361:20041722}}

@article{Read2009,
  author={Read, J. S. and Lackey, B. D. and Owen, B. J. and Friedman, J. L.}, title={Constraints on a Phenomenologically Parameterized Neutron-Star Equation of State},
  journal={Physical Review D}, volume={79}, pages={124032}, year={2009}, doi={10.1103/PhysRevD.79.124032}}

@article{LattimerPrakash2016,
  author={Lattimer, J. M. and Prakash, M.}, title={The Equation of State of Hot, Dense Matter and Neutron Stars},
  journal={Physics Reports}, volume={621}, pages={127--164}, year={2016}, doi={10.1016/j.physrep.2015.12.005}}

\end{document}